\documentclass[12pt,a4paper]{article}
\usepackage{graphics,graphicx}
\usepackage{amssymb,epsfig,amsmath,euscript,array}
\usepackage{cite}
\makeatletter
\@addtoreset{equation}{section}
\makeatother
\renewcommand{\theequation}{\thesection.\arabic{equation}}

\newcommand{\startappendix}{
\setcounter{section}{0}
\renewcommand{\thesection}{\Alph{section}}
\renewcommand{\theequation}{\Alph{section}.\arabic{equation}}}
\newcommand{\Appendix}[1]{
\refstepcounter{section}
\begin{flushleft}
{\Large\bf Appendix \thesection: #1}
\end{flushleft}}

\newcounter{multieqs}

\newcommand{\be}{\begin{equation}}
\newcommand{\ee}{\end{equation}}

\newcommand{\bm}[1]{\mbox{\boldmath $#1$}}

\newcommand{\kslash}{k \!\!\! / }

\newcommand{\lslash}{l \!\! / }
\newcommand{\Pslash}{P \!\!\!\! / }

\newcommand{\islash}{i \!\!\! / }
\newcommand{\jslash}{j \!\!\! / }
\newcommand{\aslash}{a \!\!\! / }
\newcommand{\bslash}{{b \hspace{-6pt} \slash} }

\newcommand{\onslash}{1 \!\!\! / }
\newcommand{\twslash}{2 \!\!\!/ }
\newcommand{\thslash}{3 \!\!\!/ }
\newcommand{\foslash}{4 \!\!\! / }
\newcommand{\fislash}{5 \!\!\! / }

\newcommand{\mslash}{m \!\!\! / }

\def\bd{\begin{document}}
\def\ed{\end{document}}
\def\nn{\nonumber}
\def\bea{\begin{eqnarray}}
\def\eea{\end{eqnarray}}
\def\eps{\epsilon}

\def\ab{(ijab)}
\def\ba{(ijba)}
\def\ijab{{\tr}_{+}(\islash\, \jslash\, \aslash \, \bslash)}
\def\ijba{{\tr}_{+}(\islash\, \jslash\, \bslash \, \aslash)}
\def\ijaP{{\tr}_{+}(\islash\, \jslash\, \aslash \, \Pslash)}
\def\ijPLa{{\tr}_{+}(\islash\, \jslash\, \Pslash_L \, \aslash)}
\def\ijaPL{{\tr}_{+}(\islash\, \jslash\, \aslash \, \Pslash_L)}
\def\ijPLza{{\tr}_{+}(\islash\, \jslash\, \Pslash_{L;z} \, \aslash)}
\def\ijaPLz{{\tr}_{+}(\islash\, \jslash\, \aslash \, \Pslash_{L;z})}
\def\ijPa{{\tr}_{+}(\islash\, \jslash\, \Pslash \, \aslash)}
\def\iaPb{{\tr}_{+}(\islash\, \aslash\, \Pslash \, \bslash)}
\def\ibPa{{\tr}_{+}(\islash\, \bslash\, \Pslash \, \aslash)}
\def\ijPmu{{\tr}_{+}(\islash\, \jslash\, \Pslash \, \mu)}
\def\ibmuP{{\tr}_{+}(\islash\, \bslash\, \mu \, \Pslash)}
\def\ibmua{{\tr}_{+}(\islash\, \bslash\, \mu \, \aslash)}
\def\iamub{{\tr}_{+}(\islash\, \aslash\, \mu \, \bslash)}
\def\jaPb{{\tr}_{+}(\jslash\, \aslash\, \Pslash \, \bslash)}
\def\ijmuP{{\tr}_{+}(\islash\, \jslash\, \mu \, \Pslash)}
\def\ijmum{{\tr}_{+}(\islash\, \jslash\, \mu \, \mslash)}
\def\ijmmu{{\tr}_{+}(\islash\, \jslash\, \mslash \, \mu)}
\def\ijmP{{\tr}_{+}(\islash\, \jslash\, \mslash \, \Pslash)}
\def\iabP{{\tr}_{+}(\islash\, \aslash\, \bslash \, \Pslash)}
\def\ijbP{{\tr}_{+}(\islash\, \jslash\, \bslash \, \Pslash)}
\def\jbPa{{\tr}_{+}(\jslash\, \bslash\, \Pslash \, \aslash)}
\def\ijPb{{\tr}_{+}(\islash\, \jslash\, \Pslash \, \bslash)}
\def\jbmua{{\tr}_{+}(\jslash\, \bslash\, \mu \, \aslash)}

\def\loablt{ {\tr}_{+}(\lslash_1\, \aslash \, \bslash\, \lslash_2)}

\def\ijlolt{{\tr}_{+}(\islash\, \jslash\, \lslash_1 \, \lslash_2)}
\def\ijltlo{{\tr}_{+}(\islash\, \jslash\, \lslash_2 \, \lslash_1)}
\def\ibloa{{\tr}_{+}(\islash\, \bslash\, \lslash_1 \, \aslash)}
\def\jaltb{{\tr}_{+}(\jslash\, \aslash\, \lslash_2 \, \bslash)}
\def\ialtb{{\tr}_{+}(\islash\, \aslash\, \lslash_2 \, \bslash)}
\def\bltloa{{\tr}_{+}(\bslash\, \lslash_2\, \lslash_1 \, \aslash)}
\def\jbloa{{\tr}_{+}(\jslash\, \bslash\, \lslash_1 \, \aslash)}
\def\ibPb{{\tr}_{+}(\islash\, \bslash\, \Pslash \, \bslash)}
\def\ijltb{{\tr}_{+}(\islash\, \jslash\, \lslash_2 \, \bslash)}

\def\ijloa{{\tr}_{+}(\islash\, \jslash\,  \lslash_1 \, \aslash)}
\def\ijblt{{\tr}_{+}(\islash\, \jslash\,  \bslash \, \lslash_2)}

\def\jakb{{\tr}_{+}(\jslash\, \aslash\, \kslash \, \bslash)}
\def\iakb{{\tr}_{+}(\islash\, \aslash\, \kslash \, \bslash)}

\def\tofo{{\tr}_{+}(\onslash\, \thslash\, \twslash \, \foslash)}
\def\foto{{\tr}_{+}(\onslash\, \thslash\, \foslash \, \twslash)}
\def\tofi{{\tr}_{+}(\onslash\, \thslash\, \twslash \, \fislash)}
\def\fito{{\tr}_{+}(\onslash\, \thslash\, \fislash \, \twslash)}

\def\lrangle#1#2{\langle #1\,#2\rangle}

\def\Li{{$\rm Li}_2$}

\let\bm=\bibitem
\let\la=\label

\def\npb#1#2#3{Nucl. Phys. {\bf{B#1}} #3 (#2)}
\def\plb#1#2#3{Phys. Lett. {\bf{#1B}} #3 (#2)}
\def\prl#1#2#3{Phys. Rev. Lett. {\bf{#1}} #3 (#2)}
\def\prd#1#2#3{Phys. Rev. {D \bf{#1}} #3 (#2)}
\def\cmp#1#2#3{Comm. Math. Phys. {\bf{#1}} #3 (#2)}
\def\cqg#1#2#3{Class. Quantum Grav. {\bf{#1}} #3 (#2)}
\def\nppsa#1#2#3{Nucl. Phys. B (Proc. Suppl.) {\bf{#1A}}#3 (#2)}
\def\ap#1#2#3{Ann. of Phys. {\bf{#1}} #3 (#2)}
\def\ijmp#1#2#3{Int. J. Mod. Phys. {\bf{A#1}} #3 (#2)}
\def\rmp#1#2#3{Rev. Mod. Phys. {\bf{#1}} #3 (#2)}
\def\mpla#1#2#3{Mod. Phys. Lett. {\bf A#1} #3 (#2)}
\def\jhep#1#2#3{J. High Energy Phys. {\bf #1} #3 (#2)}
\def\atmp#1#2#3{Adv. Theor. Math. Phys. {\bf #1} #3 (#2)}
\newcommand{\EQ}[1]{\begin{equation} #1 \end{equation}}
\newcommand{\AL}[1]{\begin{subequations}\begin{align} #1 \end{align}\end{subequations}}
\newcommand{\SP}[1]{\begin{equation}\begin{split} #1 \end{split}\end{equation}}
\newcommand{\ALAT}[2]{\begin{subequations}\begin{alignat}{#1} #2 \end{alignat}
                        \end{subequations}}
\def\beqa{\begin{eqnarray}}
\def\eeqa{\end{eqnarray}}
\def\beq{\begin{equation}}
\def\eeq{\end{equation}}
\def\sst{\scriptscriptstyle}
\def\thetabar{\bar\theta}
\def\Tr{{\rm Tr}}
\def\one{\mbox{1 \kern-.59em {\rm l}}}
 \def\Nh{\hat{N}}

\def\a{\alpha}      \def\da{{\dot\alpha}}
\def\b{\beta}       \def\db{{\dot\beta}}
\def\g{\gamma}  \def\G{\Gamma}  \def\cdt{\dot\gamma}
\def\d{\delta}  \def\D{\Delta}  \def\ddt{\dot\delta}
\def\e{\epsilon}        \def\vare{\varepsilon}
\def\f{\phi}    \def\F{\Phi}    \def\vvf{\f}
\def\h{\eta}
\def\k{\kappa}
\def\l{\lambda} \def\L{\Lambda}
\def\m{\mu} \def\n{\nu}
\def\o{\omega}
\def\p{\pi} \def\P{\Pi}
\def\r{\rho}
\def\s{\sigma}  \def\S{\Sigma}
\def\t{\tau}
\def\th{\theta} \def\Th{\Theta} \def\vth{\vartheta}
\def\X{\Xeta}
\def\z{\zeta}
\def\cA{{\cal A}} \def\cB{{\cal B}} \def\cC{{\cal C}}
\def\cD{{\cal D}} \def\cE{{\cal E}} \def\cF{{\cal F}}
\def\cG{{\cal G}} \def\cH{{\cal H}} \def\cI{{\cal I}}
\def\cJ{{\cal J}} \def\cK{{\cal K}} \def\cL{{\cal L}}
\def\cM{{\cal M}} \def\cN{{\cal N}} \def\cO{{\cal O}}
\def\cP{{\cal P}} \def\cQ{{\cal Q}} \def\cR{{\cal R}}
\def\cS{{\cal S}} \def\cT{{\cal T}} \def\cU{{\cal U}}
\def\cV{{\cal V}} \def\cW{{\cal W}} \def\cX{{\cal X}}
\def\cY{{\cal Y}} \def\cZ{{\cal Z}}
\def\ua{\underline{\alpha}}
\def\ub{\underline{\phantom{\alpha}}\!\!\!\beta}
\def\uc{\underline{\phantom{\alpha}}\!\!\!\gamma}
\def\um{\underline{\mu}}
\def\ud{\underline\delta}
\def\ue{\underline\epsilon}
\def\una{\underline a}\def\unA{\underline A}
\def\unb{\underline b}\def\unB{\underline B}
\def\unc{\underline c}\def\unC{\underline C}
\def\und{\underline d}\def\unD{\underline D}
\def\une{\underline e}\def\unE{\underline E}
\def\unf{\underline{\phantom{e}}\!\!\!\! f}\def\unF{\underline F}
\def\unm{\underline m}\def\unM{\underline M}
\def\unn{\underline n}\def\unN{\underline N}
\def\unp{\underline{\phantom{a}}\!\!\! p}\def\unP{\underline P}
\def\unq{\underline{\phantom{a}}\!\!\! q}
\def\unQ{\underline{\phantom{A}}\!\!\!\! Q}
\def\unH{\underline{H}}
\def\As {{A \hspace{-6.4pt} \slash}\;}
\def\bs {{b \hspace{-6.4pt} \slash}\;}
\def\Ds {{D \hspace{-6.4pt} \slash}\;}
\def\ds {{\del \hspace{-6.4pt} \slash}\;}
\def\ss {{\s \hspace{-6.4pt} \slash}\;}
\def\ks {{ k \hspace{-6.4pt} \slash}\;}
\def\ps {{p \hspace{-6.4pt} \slash}\;}
\def\pas {{{p_1} \hspace{-6.4pt} \slash}\;}
\def\pbs {{{p_2} \hspace{-6.4pt} \slash}\;}
\def\Ps {{P \hspace{-6.4pt} \slash}\;}
\def\Qs {{Q \hspace{-6.4pt} \slash}\;}
\def\Fh{\hat{F}}
\def\Vh{\hat{V}}
\def\Xh{\hat{X}}
\def\ah{\hat{a}}
\def\xh{\hat{x}}
\def\yh{\hat{y}}
\def\ph{\hat{p}}
\def\xih{\hat{\xi}}
\def\psit{\tilde{\psi}}
\def\Psit{\tilde{\Psi}}
\def\tht{\tilde{\th}}
\def\lt{\tilde{\lambda}}
\def\llt{\tilde{l}}
\def\At{\tilde{A}}
\def\Qt{\tilde{Q}}
\def\Rt{\tilde{R}}
\def\Nt{\tilde{N}}

\def\at{\tilde{a}}
\def\st{\tilde{s}}
\def\ft{\tilde{f}}
\def\pt{\tilde{p}}
\def\qt{\tilde{q}}
\def\vt{\tilde{v}}
\def\nt{\tilde{n}}
\def\delb{\bar{\partial}}
\def\bz{\bar{z}}
\def\bD{\bar{D}}
\def\bB{\bar{B}}
\def\bk{{\bf k}}
\def\bl{{\bf l}}
\def\bp{{\bf p}}
\def\bq{{\bf q}}
\def\br{{\bf r}}
\def\bx{{\bf x}}
\def\by{{\bf y}}
\def\bR{{\bf R}}
\def\bV{{\bf V}}
\def\d{\delta}\def\D{\Delta}\def\ddt{\dot\delta}
\def\pa{\partial} \def\del{\partial}
\def\xx{\times}
\def\uno{\mbox{1 \kern-.59em {\rm l}}}
\def\trp{^{\top}}
\def\inv{^{-1}}
\def\dag{{^{\dagger}}}
\def\pr{^{\prime}}
\def\lan{\langle}
\def\ran{\rangle}
\def\rar{\rightarrow}
\def\lar{\leftarrow}
\def\lrar{\leftrightarrow}
\newcommand{\0}{\,\!}      %this is just NOTHING!
\def\one{1\!\!1\,\,}
\def\im{\imath}
\def\jm{\jmath}
\newcommand{\tr}{\mbox{tr}}
\newcommand{\slsh}[1]{/ \!\!\!\! #1}
\def\vac{|0\rangle}
\def\lvac{\langle 0|}
\def\hlf{\frac{1}{2}}
\def\ove#1{\frac{1}{#1}}
\def\Box{\square}
\def\ZZ{\mathbb{Z}}
\def\CC#1{({\bf #1})}
\def\bcomment#1{}
\def\bfhat#1{{\bf \hat{#1}}}
\def\VEV#1{\left\langle #1\right\rangle}
\newcommand{\ex}[1]{{\rm e}^{#1}} \def\ii{{\rm i}}
\def\rr{{\rm r}} \def\rs{{\rm s}}\def\rv{{\rm v}}
\def\ri{{\rm i}}\def\rj{{\rm j}}
\newcommand{\lrbrk}[1]{\left(#1\right)}
\newcommand{\sfrac}[2]{{\textstyle\frac{#1}{#2}}}

\def\Li{{\rm Li}_2}

\def\Li2{{\rm Li}_2}

\font\mybb=msbm10 at 12pt
\def\bb#1{\hbox{\mybb#1}}

\font\myBB=msbm10 at 18pt
\def\BB#1{\hbox{\myBB#1}}

\begin{document}

\begin{flushright}
hep-th/0510253\\
QMUL-PH-05-12
\end{flushright}

\vspace{20pt}

\begin{center}

{\Large \bf From Trees to Loops and Back  }
%\vspace{0.3cm}
%
\vspace{33pt}

{\bf {\mbox{Andreas  Brandhuber,  Bill Spence and Gabriele  Travaglini}}}%
\footnote{{\sffamily \{\tt a.brandhuber, w.j.spence, g.travaglini\}@qmul.ac.uk }}

{\em Department of Physics\\
Queen Mary, University of
London\\
Mile End Road, London, E1 4NS\\
United Kingdom}
\vspace{40pt}

{\bf Abstract}

\end{center}

\noindent
We argue that generic one-loop scattering amplitudes
in supersymmetric Yang-Mills theories can 
be computed equivalently with MHV diagrams or with
Feynman diagrams. We first present 
a general proof of the covariance of one-loop non-MHV  
amplitudes obtained from MHV diagrams.
This proof relies only on
the local character in Minkowski space of MHV vertices  
and on an application of the Feynman Tree Theorem. 
We then show that the discontinuities 
of one-loop scattering amplitudes computed with MHV diagrams 
are precisely the same as  those computed with standard methods. 
Furthermore, we analyse collinear limits and soft limits 
of generic non-MHV amplitudes in supersymmetric Yang-Mills theories
with one-loop MHV diagrams. 
In particular, we find a simple explicit derivation of 
the universal one-loop splitting functions 
in supersymmetric Yang-Mills theories 
to all orders in the dimensional regularisation parameter, 
which is in complete agreement with known results. 
Finally, we present concrete and illustrative applications 
of Feynman's Tree Theorem to one-loop MHV diagrams
as well as to one-loop Feynman diagrams.

\vspace{0.5cm}

\setcounter{page}{0}
\thispagestyle{empty}
\newpage

%%%%%%%%%%%%%%%%%%%%%%%%%%%%%%%%%%%%%%%%%%%%%%%%%%%%%%%%%%%%%%%%

                     \section{Introduction}

%%%%%%%%%%%%%%%%%%%%%%%%%%%%%%%%%%%%%%%%%%%%%%%%%%%%%%%%%%%%%%%%

\setcounter{footnote}{0}

Following the seminal paper \cite{witten},
Cachazo, Svr\v{c}ek and Witten
proposed in \cite{csw} a novel method for
calculating generic tree-level
scattering amplitudes of gluons.
This method  makes use of MHV amplitudes
continued off-shell in an appropriate way as
vertices of a new perturbative expansion
of Yang-Mills theories, and offers a powerful
diagrammatic alternative to Feynman diagrams.
In a sense, this new expansion
is close in spirit to the S-matrix approach
\cite{bible}, as its building blocks are scattering amplitudes.
The similarities with the S-matrix approach persist
when loop amplitudes are considered -- for example,
it was shown in \cite{bst}
that the infinite sequence of
one-loop amplitudes with the MHV configuration
in $\cN =4$ super Yang-Mills can be written as dispersion
integrals, whose explicit evaluation leads to precise agreement
with the previously known expressions obtained by
Bern, Dixon, Dunbar and Kosower in \cite{bdk1} using the
unitarity-based approach.

At tree level, numerous successful applications of the MHV diagram
method have been carried out so far
\cite{Zhu}--\!\!\cite{Ozeren:2005mp}.
An elegant proof of the method
at tree level was presented in \cite{bcfw} based on
the analytic properties of the scattering amplitudes.
The same paper also discussed the proof of a
new recursion relation \cite{bcf}
which allows one to calculate scattering
amplitudes at tree level very efficiently.
It is now clear that the existence of recursion relations
is by no means limited to tree amplitudes, nor to amplitudes
involving only massless particles.
Indeed, important extensions of the BCFW recursion relation
to loop amplitudes in massless gauge theories
have been found in
\cite{bdk-rational1,bdk-rational2, bdk-rational3, coeffrec}.
Moreover, in \cite{snvp,Badger:2005jv}
new recursion relations were derived for
amplitudes involving massless and massive particles.
An interesting application of the
MHV diagram method involving a massive Higgs and top quark
had also been studied earlier in  \cite{lnv}.
Very recently, the paper \cite{risager}
established an explicit connection between MHV diagrams
and the BCFW recursion relation for
tree-level amplitudes \cite{bcf,bcfw}.
This connection, together with the existence of
recursion relations for scattering amplitudes of
gravitons \cite{bbst3,cs}, was later exploited in \cite{swan}
and led to a derivation of
new rules for calculating scattering amplitudes
in  General Relativity.

At one loop, the MHV diagram method has
been applied to re-derive the MHV scattering amplitudes
in $\cN=1$ super Yang-Mills \cite{qr,bbst},
and also to derive new results -- specifically
the cut-constructible part of the MHV amplitude
in pure Yang-Mills \cite{bbst2}
with negative helicity gluons in arbitrary positions,
which generalises earlier calculations of
\cite{Bern:1994cg, bdk9302280}.
Although these are certainly very strong tests of the MHV method,
so far no proof has been presented that
(cut-constructible parts of) one-loop
scattering amplitudes computed
with MHV diagrams agree 
with those obtained using conventional methods.
In this paper, we will give very strong evidence 
that this is indeed true.

A key step in showing the equivalence of the
two approaches
is a proof of the
covariance of the one-loop amplitudes
calculated in the MHV diagram approach.
With MHV diagrams one introduces an arbitrary
reference null momentum $\eta$ in order to define the off-shell
continuation of amplitudes \cite{csw}. 
Clearly, it is crucial to show that
physical scattering amplitudes are covariant --
i.e.~independent of the particular choice made for $\eta$.
It was proved in \cite{csw} that,
at tree level, the result of a calculation based on
MHV diagrams is indeed independent
of the choice of the reference spinor
after summing over all MHV diagrams.
At one loop, explicit calculations
\cite{bst,qr,bbst,bbst2} of MHV one-loop amplitudes in
Yang-Mills theories show  that
the results are $\eta$-independent.
However, the cancellation of $\eta$-dependent terms is
rather non-trivial, and is achieved in general
by combining terms which originate from different
MHV diagrams. In addition, it was necessary
in some cases to resort to numerical methods.

The analytic proof of covariance we will present
in this paper makes use of a simple and beautiful result
in field theory due to Feynman, known as the
Feynman Tree Theorem \cite{F1,F2,F3}.
Anticipating our story a little, we would like to
mention here that the application of this theorem leads to
an alternative way of calculating loop amplitudes
(by no means limited to the one-loop level or
to massless particles), which
are then expressed as sums of terms with one or more loop lines
cut open by delta functions -- therefore Feynman's theorem
allows one to calculate loops from trees.

We will review in detail the Feynman Tree Theorem
in section 2, but we would like to stress
two important facts that are at the core of its derivation:
\begin{itemize}
\item[{\bf 1.}]
A key observation is that
a Feynman propagator can be decomposed into a retarded
(or advanced) propagator and a delta function term
which is supported on shell. These delta functions
have the effect, mentioned earlier, of cutting open internal
loop legs.
\item[{\bf 2.}]
The locality of the interaction is crucial
for the applicability of the theorem, in that it
guarantees that a loop amplitude calculated with
Feynman propagators replaced by retarded (or advanced)
propagators actually vanishes. As we shall see in
section 2, this requirement is central to
proving Feynman's theorem.
\end{itemize}
In the MHV diagram approach, MHV vertices are
connected simply by scalar Feynman propagators, for which
we will employ the above mentioned decomposition
into a retarded/advanced propagator plus
a term supported on shell.
The locality requirement is furthermore satisfied
thanks to the fact that an MHV tree amplitude
can be thought of as a local interaction in Minkowski space
\cite{csw}. This important result stems from the fact that
an MHV amplitude at tree level
is localised on a complex line in twistor space
\cite{witten}, together with the incidence relation of
twistor theory, which establishes a
correspondence between lines in twistor space and  points
in Minkowski space.
We would also like to mention that this decomposition
of a Feynman propagator
has already been used in \cite{csw2}.
Specifically, the analysis of \cite{csw} shows that
tree-level scattering amplitudes in Yang-Mills have support
on unions of lines in twistor space;
Feynman's decomposition was then used in
\cite{csw2} in order to prove that
the lines pairwise intersect.%
\footnote{This conclusion is reached if one discards
the delta function terms in the propagator decomposition.
At tree level, these delta functions would contribute only
at exceptional configurations for the external momenta.}

We establish the covariance of one-loop amplitudes
in the MHV diagram approach in section 3, where we discuss
a few examples which  illustrate our strategy,
and which can immediately be generalised to one-loop amplitudes
with arbitrary helicity configurations.
We will then argue in section 4 that these one-loop amplitudes
have the same physical discontinuities and poles
as the amplitudes derived using Feynman diagrams.
That MHV diagrams have the same discontinuities
as Feynman diagrams is simple to show. As far as
poles are concerned, we will present
a detailed derivation of the collinear limits 
of amplitudes in the MHV diagrams approach, 
and  will also discuss soft limits.

Collinear limits have a universal structure,
encoded into splitting functions,
which emerges neatly in the MHV diagram approach
considered in this paper.
The result of our analysis of collinear limits
agrees with the well-known
supersymmetric results of
\cite{bdk1,Bern:1994cg,Bern:1995ix,david,ku,vittorio,vittorio2}.%
\footnote{In non-supersymmetric theories
we obtain the cut-constructible
part of the amplitude, missing certain rational terms.}
Specifically, we reproduce the very simple
expression of \cite{vittorio}
for the one-loop gluon splitting function
valid to all orders in $\e$ (the dimensional regularisation
parameter),  given in terms of 
hypergeometric functions (this expression is
also identical to that found in
\cite{ku}).
We establish our result using a new form of the
all-order in $\e$ two-mass easy box function,
presented in section 4 (and further discussed in 
the Appendix), whose $\e\to 0$ limit
was given in (7.1) of \cite{bst}.
As discussed in section 5 of that paper,
this form of the box function
has a simpler analytic continuation to the
physical region than the usual expression.

One should also make sure that unphysical,
$\eta$-dependent terms cannot appear which 
would spoil the matching of discontinuities 
and poles between the MHV diagrams result and
the Feynman diagrams calculation.
This fact is central if one is to prove 
the equivalence of MHV and Feynman diagrams, 
and is indeed guaranteed by our proof of covariance
at one loop presented in section 3.

We conclude the paper by giving in section 5
two simple and instructive applications
of the Feynman Tree Theorem to concrete calculations:
that of a bubble diagram in a generic local theory,
and the calculation of an MHV amplitude at one loop
using MHV diagrams.
In this second example, particularly relevant
for this paper, we will present a re-derivation of
the one-loop integration measure of \cite{bst}
based on the application of the Feynman Tree Theorem.
Finally, we summarise our results in section 6. 
An Appendix contains various forms of the two-mass easy
box function to all orders in $\e$.

Further work on one- and multi-loop
amplitudes appears in
\cite{csw3}--\!\!\cite{Forde:2005hh}
and \cite{babis1}--\!\!\cite{Buc-Cac}.

%%%%%%%%%%%%%%%%%%%%%%%%%%%%%%%%%%%%%%%%%%%%%%%%%%%%%%%%%%%%%%%%

\section{The Feynman Tree Theorem}

%%%%%%%%%%%%%%%%%%%%%%%%%%%%%%%%%%%%%%%%%%%%%%%%%%%%%%%%%%%%%%%%

The main ingredient of the Feynman Tree Theorem
is a decomposition of the Feynman propagator into
a retarded (or advanced) propagator and
an additional contribution which is localised on the
mass shell. To see this, let us recall some simple
but important facts about propagators.

Consider a free scalar field
(which is not necessarily massless).
In momentum space, the Feynman propagator, $\D_F$,
and the retarded and advanced Green functions,
$\D_R$ and $\D_A$, are given by
\beqa
\label{ofey}
\Delta_F (P) & := & {i \over 2 \o} \Big[ {1 \over P_0 - \o + i \vare} -
{1 \over P_0 + \o - i \vare} \Bigr] \ = \
 {i \over P_0^2 - \o^2 + i \vare}
\ ,
\\ \cr
\label{ret}
\Delta_R (P) & := & {i \over 2 \o} \Big[ {1 \over P_0 - \o + i \vare} -
{1 \over P_0 + \o + i \vare} \Bigr] \ = \
 {i \over (P_0 + i  \vare)^2  - \o^2}
\ ,
\\ \cr
\label{adv}
\Delta_A (P) & := & {i \over 2 \o} \Big[ {1 \over P_0 - \o - i \vare} -
{1 \over P_0 + \o - i \vare} \Bigr] \ = \
 {i \over (P_0 - i  \vare)^2  - \o^2}
\ ,
\eeqa
where $\o := \sqrt{ |\vec{P}|^2 + m^2}$ and $\vare \to 0^+$.
We can immediately re-cast \eqref{ofey}-\eqref{adv} as
\beqa
\label{ofey2}
\Delta_F (P) & = &
 {i \over P^2 - m^2 + i \vare}
\ ,
\\ \cr
\label{ret2}
\Delta_R (P) & = &
 {i \over P^2 - m^2 + i \vare \, {\rm sgn}(P_0)}
\ ,
\\ \cr
\label{adv2}
\Delta_A (P) & = &
 {i \over P^2 - m^2 - i \vare \, {\rm sgn} (P_0)}
\ ,
\eeqa
where $P^2:= P_\m P^\m = P_0^2 - |\vec{P}|^2 $.

Of course, it is the Feynman propagator \eqref{ofey} that enters
the calculation of scattering amplitudes.
In \cite{F1,F2} Feynman made use of the simple fact that
the retarded and the advanced propagator differ
from the Feynman propagator only by a delta-function contribution
localised on the mass-shell $P^2 = m^{2}$, namely
\beqa
\label{decret}
\Delta_R (P) & = & \Delta_F (P) \, -\, 2 \pi \delta (P^2-m^{2})
\theta (-P_0)
\ ,
\\ \cr
\label{decadv}
\Delta_A (P) & = & \Delta_F (P) \, -\, 2 \pi \delta (P^2-m^{2})
\theta (P_0)
\ .
\eeqa
This can be seen by comparing \eqref{ret} and  \eqref{adv}
to \eqref{ofey}, and using the identity
$ 1/ (x+i \vare) = P (1/x) - i \pi \d (x)$, where $P$ stands
for the principal value prescription.

Now, imagine we wish to compute a generic one-loop amplitude
in a locally interacting theory.
In momentum space, this can formally be written as%
\footnote{In \eqref{oneloop} we omit a delta function
which enforces momentum conservation,
$\sum_{i} K_i = 0$.
Furthermore, the use of a regulator for potential ultraviolet
and infrared divergences in loop integrals
will be understood in the rest
of the paper.
}
\beq
\label{oneloop}
\cL \ = \ \int\! {d^4L \over (2 \pi)^4}
\, f (L, \{ K_i\} ) \, \prod_{i} \Delta_F (L+K_i)
\ .
\eeq
Here $K_i$ is a sum of external momenta
(for the case of colour-ordered amplitudes to be considered later,
the $K_{i}$ are sums of cyclically adjacent momenta), and
$f$ is a polynomial function of the
loop momentum and of the external momenta
produced by the numerators of
particle propagators and interaction vertices.
Following Feynman, we consider now the quantity
\beq
\label{zero}
\cL_R \ := \ \int\! {d^4L \over (2 \pi)^4}
\, f (L, \{K_i\}) \, \prod_{i} \Delta_R (L+K_i)
\ ,
\eeq
which is obtained from \eqref{oneloop}
by simply replacing all Feynman propagators
with  retarded propagators.
Crucially,%
\footnote{The use of advanced propagators
instead of retarded propagators
would lead to the same conclusion.}
\beq
\label{zero2}
\cL_R \ = \ 0
\ .
\eeq
This result can be proved in two ways \cite{F1,F2}.
Firstly, one can work directly in momentum space and note
that all the poles in \eqref{zero} lie below the real $L_0$ axis;
thus the integration contour can be closed by a large semicircle
in the upper complex $L_0$ plane, and \eqref{zero2}
follows immediately.
Alternatively, one can work in configuration space, where a generic
loop amplitude is expressed as
\beq
\label{loop-config}
\cL \ = \ \int\!\!\prod_i  d^4x_i \
\D_F (x_1-x_2) \cV(x_2) \D_F (x_2 - x_3) \cV( x_3)
\cdots \D_F (x_n - x_1) \cV (x_1)
\ ,
\eeq
where
\beq
\D_F (x) \ := \
\int\!{d^4P \over (2 \pi)^4} \ e^{-i P x}\, \D_F (P)
\ ,
\eeq
and $\cV (x)$ stands for an arbitrary local interaction.
By replacing all Feynman propagators in
\eqref{loop-config} with retarded (or advanced) propagators,
one obtains
\beq
\label{loop-config2}
\cL_R \ = \ \int\!\prod_i  d^4x_i \
\D_R (x_1-x_2) \cV(x_2) \D_R (x_2 - x_3) \cV( x_3)
\cdots  \D_R (x_n - x_1) \cV (x_1)
 \ = \ 0
\ .
\eeq
This follows from the fact that the integration in
$\cL_R$ has support for
\beq
\label{ctl}
t_{1} > t_{2} > \cdots > t_{n}> t_{1}
\ ,
\eeq
and, since there are no closed time-like curves
in Minkowski space,
$\mathcal{L}_R$ must vanish.

Note that the time ordering in \eqref{ctl}
is a consequence of the fact  that
the retarded (advanced) propagator
$\D_R (x)$ ($\D_A (x)$) has support
only inside the future (past) lightcone.
This is immediately seen by looking at their
expressions in configuration space.
Performing  first the $P_0$ integration, we see that
for $t < 0$  ($t > 0$) the integration contour
can be closed with a large semicircle
in the upper complex plane, with no singularity
being enclosed by the contour.
The result of the integration,
and hence the retarded (advanced) propagator
vanishes  unless   $t > 0$  ($t < 0$).
Since the retarded (advanced) propagator is Lorentz invariant,
one concludes that it has support
inside the future (past) lightcone --
the retarded and advanced propagators
are thus causal propagators.
For the case of massless particles,
$P^2 =0$, one finds very simple
explicit expressions for these propagators,%
\footnote{The retarded and advanced propagators
are related by $\D_R( x) = \D_A (-x)$.}
\beqa
\Delta_R (x) & = & {1 \over 2\pi i} \theta (t) \delta (x^2)
\ ,
\\ \cr
\Delta_A (x) & = & {1\over 2\pi i} \theta (-t) \delta (x^2)
\ .
\eeqa
Hence, in the massless case the retarded (advanced)
propagators have support on the future (past) lightcone.

We now insert the decomposition
\eqref{decret} into \eqref{zero}, and using
\eqref{zero2} we obtain
\beq
\label{zero3}
\cL_R \ := \ \int\! {d^4L \over (2 \pi)^4}
\, f (L, \{K_i\}) \, \prod_{i}
\Big[
\Delta_F (L+K_i) - 2 \pi \d^{(-)} \big( ( L+K_i)^2 \big)
\Big] \ = \ 0
\ .
\eeq
Expanding the product in \eqref{zero3}
we arrive at the result \cite{F1,F2}
\beq
\label{ftt}
\cL  \ = \
- \, \int\! {d^4L \over (2 \pi)^4}
\, f (L, \{K_i\}) \, \prod_{i}^{\ \ \ \ \ \prime}
\Big[
\Delta_F (L+K_i) - 2 \pi \d^{(-)} \big( ( L+K_i)^2 \big)
\Big]
\ ,
\eeq
where
$ \delta^{\pm} (L^2) : = \theta ( \pm L_0 ) \delta (L^2)$,
and the prime on the product means that we only keep terms
with at least one delta function.
The left hand side of \eqref{ftt} is the term in the expansion of
\eqref{zero3} that contains only
Feynman propagators and no delta function,
and is precisely equal to the
physical one-loop amplitude \eqref{oneloop}
which we wish to calculate.%
\footnote{
\label{foo}
Notice that if more than four delta functions appear
on the right-hand side, then
the corresponding term will in general vanish.
On the other hand, four delta functions freeze
the loop integral, as in \cite{bcf-gen}.}

Eq.~\eqref{ftt} is the Feynman Tree Theorem.
We can also re-cast it in a more transparent form,
\beq
\label{ftt2}
\cL \ = \ \cL_{\rm 1-cut}\, + \,
\cL_{\rm 2-cut}\, + \, \cL_{\rm 3-cut}\, +  \, \cL_{\rm 4-cut}
\ ,
\eeq
where $\cL_{\rm {\it p}-cut}$ is the sum of all
the terms in \eqref{ftt} with precisely $p$ delta functions.
Each delta function cuts open an internal loop leg,
and therefore a term with $p$ delta functions computes a
$p$-particle cut in a kinematical channel determined
by the cut propagators (whose momentum
is set on shell by the delta functions).%
\footnote{Note that the cuts
appearing in the Feynman tree theorem are
not identical to the conventional
\cite{Landau:1959fi,Cutkosky:1960sp} unitarity cuts.
This is a consequence of the presence of a
step function multiplying the delta function
in the decomposition of the Feynman propagator \eqref{decret}
and \eqref{decadv}.
A detailed discussion of this subtlety can be found
in section 5.1.3.}

Feynman's Tree Theorem \eqref{ftt2} states
that a one-loop diagram
can be expressed as a sum over
all possible cuts of the loop diagram.
By iteration this statement can also be applied
to higher-loop diagrams.
At one loop, a Feynman $p$-particle cut decomposes
the diagram into $p$ tree diagrams --
in essence, the Tree Theorem allows us to calculate
loops from trees.
The process of cutting puts internal lines
on shell; of course there are remaining
phase space integrations to be performed, but these are
generically easier than the original loop integration.
Moreover, this also implies that one-loop
diagrams and scattering amplitudes
can be determined from on-shell data alone.

As discussed in the Introduction,
the proof of the Feynman Tree Theorem relies crucially on the
locality of the interaction vertices.
Had the interaction been non-local,
the causality argument used to prove
\eqref{zero3} would not hold.
In  \cite{witten} it was shown that an MHV tree-level
scattering amplitude localises on a line in twistor space.
By virtue of the incidence relation of twistor theory,
a line in twistor space corresponds to
a point in Minkowski space.
Hence, this implies that
an MHV tree amplitude can be thought of as a
local interaction in Minkowski space
\cite{csw}.

The local nature of MHV amplitudes was instrumental
in deriving the new diagrammatic rules of \cite{csw}, where
MHV amplitudes are promoted to effective vertices.
The locality of MHV vertices will also be crucial for our
application of the Feynman Tree Theorem to one-loop MHV diagrams
(rather than Feynman diagrams), which will allow us to prove
the covariance of the MHV diagram method at one loop.
This result will then be used to give 
strong evidence that MHV diagrams
at one loop give results in complete agreement with
a standard calculation based on Feynman diagrams.

Finally, we stress that the
Feynman Tree Theorem does not require the particles
to be massless, and works for supersymmetric and
non-supersymmetric theories alike.
Thus we expect that this theorem will have many other applications
that go beyond the MHV diagram method discussed here.

%%%%%%%%%%%%%%%%%%%%%%%%%%%%%%%%%%%%%%%%%%%%%%%%%%%%%%%%%%%%%

    \section{Covariance of the amplitudes at one loop}

%%%%%%%%%%%%%%%%%%%%%%%%%%%%%%%%%%%%%%%%%%%%%%%%%%%%%%%%%%%%%

In this section we will show that the sum of one-loop MHV diagrams
contributing to a one-loop scattering amplitude
is Lorentz covariant,
i.e.~does not depend on the choice of the
reference null momentum $\eta$ introduced
in order to define the off-shell continuation
of the MHV scattering amplitudes \cite{csw}.
For tree-level amplitudes, it was shown in \cite{csw}
that $\eta$-dependent terms cancel between different MHV diagrams,
and the sum of all MHV diagrams is indeed covariant.
Furthermore, it was shown in \cite{bst} that,
for MHV amplitudes at one loop, non-trivial cancellations
between contributions from different MHV one-loop diagrams
occur, leading to the correct, covariant amplitudes.
In the following we will present a new,
more general proof of the cancellation of $\eta$-dependence
which applies to MHV and non-MHV amplitudes
in supersymmetric Yang-Mills and
to the cut-constructible part of
the same amplitudes in pure Yang-Mills.

Consider the set of one-loop MHV diagrams
corresponding to a particular colour-ordered amplitude
with $q$ negative helicity gluons and $n-q$ positive
helicity gluons.%
\footnote{Cases with fermions or scalars in the
external lines can be addressed by a simple
generalisation of the line of argument
that will follow.}
Any such MHV diagram consists
of $v$ MHV vertices, where \cite{witten}
\beq
v \ = \ q\, -\, 1\, +\, l
\ ,
\eeq
and $l$ is the number of loops. Hence, at one loop, $v=q$ and as
usual these vertices are connected with
scalar Feynman propagators \cite{csw}.

Following Feynman, we now consider a different set of
one-loop MHV diagrams where, in each MHV diagram, we connect
those MHV vertices which are part
of the loop with retarded propagators
rather than Feynman propagators.
Since MHV vertices are local interactions
in Minkowski space \cite{csw}, we can apply the Feynman
Tree Theorem and each diagram in this set
gives a vanishing result.

Next, we decompose all the retarded propagators
which form the loop inside each MHV diagram
into a Feynman propagator and a delta function
contribution using \eqref{decret}.
The terms where, in each diagram, we pick the contribution
arising from the Feynman propagator are of course the
MHV diagrams corresponding to the one-loop amplitude
we are calculating.
We will now show that the sum of these MHV diagrams is covariant.

In order to do this, we observe that by the Tree Theorem
\eqref{ftt2},
the one-loop amplitude calculated using MHV diagrams
(with vertices connected
by Feynman propagators), $\cA$,
is equal to a sum of terms where at least
one internal loop line is cut open,
\beq
\label{ftt3}
\cA \ = \ \cA_{\rm 1-cut}\, + \,
\cA_{\rm 2-cut}\, + \, \cA_{\rm 3-cut}\, + \, \cA_{\rm 4-cut}
\ .
\eeq
$\cA_{\rm 1-cut}$ are one-particle cut diagrams,
i.e.~the diagrams where precisely one propagator in the loop
is replaced by a delta function. 
$\cA_{\rm 2-cut}$ are the
two-particle cut diagrams, where two
propagators are replaced by delta functions,
and so on (see footnote \ref{foo}).

The key point in the proof is that
each set of $p$-particle cut diagrams
separately sums up to a covariant expression.
The remaining phase space integrations
are also covariant, and therefore
the physical amplitude $\cA$ expressed
using the Tree Theorem as in \eqref{ftt3}
is covariant as well.

\begin{figure}[ht]
\label{Figure2}
\begin{center}
\scalebox{0.65}{\includegraphics{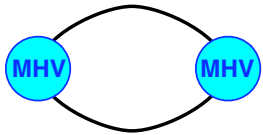}}
\end{center}
\caption{\it The MHV diagrams contributing to a
one-loop MHV scattering amplitude.
The blobs represent MHV vertices,
which should then be dressed with
external particles in all possible ways compatible
with cyclic ordering, and in such a way that the two
vertices have the MHV helicity configuration.
}
\end{figure}

Let us start illustrating this strategy
by considering the simplest case, namely
that of an MHV scattering amplitude at one loop.

\subsection{One-loop MHV amplitude}
The one-loop MHV diagrams contributing to an
$n$-point MHV amplitude are presented in Figure 1.
In our notation we only draw vertices and propagators
(or cut-propagators) connecting them. It will be understood that
we have to distribute the external gluons among the MHV vertices
in all possible ways compatible with cyclic ordering, and
the requirement that the two vertices must have
the helicity configuration of an  MHV amplitude. 
Moreover we will have to sum over all possible
helicity assignments of the internal legs and, where required,
over all possible particle species which can run in the loop
(what we are really drawing are the skeleton or quiver diagrams
of \cite{Bena:2004ry,csw2}).

\begin{figure}[ht]
\label{Figure3}
\begin{center}
\scalebox{0.65}{\includegraphics{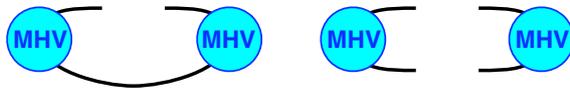}}
\end{center}
\caption{\it One-particle and two-particle MHV diagrams
contributing to the one-loop MHV scattering amplitude.
}
\end{figure}

For the MHV amplitude we have to consider only one type of
MHV diagram, represented in Figure 1,
where two MHV vertices are connected
by two Feynman propagators, which are both part of the loop.
In Figure 2 we show the one-particle and
two-particle cut diagrams which are produced in the
application of the Feynman Tree Theorem.

We start by focussing on one-particle cut diagrams.
These one-particle cut-diagrams are nothing but
tree-level diagrams, which are then integrated using a
Lorentz invariant phase space measure.
We now make the following important observation:
these tree (one-cut) diagrams would precisely sum to a
tree-level next-to-MHV (NMHV) amplitude
with $n+2$ external legs
(which would then be covariant as shown in \cite{csw}),
{\it if} we also include the set of diagrams where
the two legs into which the cut propagator is broken
are allowed to be at the same MHV vertex.
Such diagrams are obviously never generated by cutting
a loop leg in MHV diagrams of the type depicted in Figure 1.
These ``missing'' diagrams are drawn in Figure 3.
MHV rules tell us, before any phase space integration is performed,
that the combined sum of one-particle cut diagrams
of Figures 2 and 3
generates an  NMHV amplitude with $n+2$ external legs.
Since the phase
space measure is Lorentz invariant, we find that
the sum of one-particle cut diagrams,
including the missing diagrams, is covariant.

\begin{figure}[ht]
\label{Figure4}
\begin{center}
\scalebox{0.65}{\includegraphics{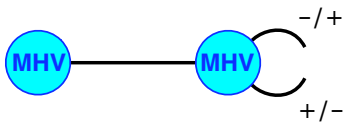}}
\end{center}
\caption{\it In this Figure we represent ``missing diagrams",
mentioned in the text.
}
\end{figure}

It remains to justify the omission
of these missing diagrams.
We will present two alternative
and somewhat complementary explanations.

The first one relies on supersymmetry.
The diagrams where two adjacent and opposite helicity
legs from the same MHV vertex are sewn together vanish
when summed over particle species in a supersymmetric theory.
Individual diagrams before summing over particle species diverge
because of the collinearity of the momenta of the two legs,
but the sum over particle species vanishes
even before integration.
So we discover that we could have actually
included these diagrams from the start,
since their contribution is zero.
Hence one-particle cut diagrams of
MHV one-loop amplitudes generate phase space integrals of
tree-level NMHV amplitudes, and are, therefore, covariant.

We now give a different explanation
which does not rely on supersymmetry.
Consider again the missing one-particle cut diagrams
of Figure 3.
Because of the delta function
which cuts open one of the internal lines,
these missing diagrams only need to be
considered in a strict (anti-)collinear limit:
if $l$ is the cut loop momentum ($l^2 = 0$),
the two legs into which this  is broken have momentum
$l_1 = l$ and $l_2 = -l$, respectively (the minus sign comes from
the fact that all our momenta are considered as outgoing).
Moreover, the two legs becoming (anti-)collinear
always have opposite helicities, since they arise from
cutting open an internal loop leg. In the anti-collinear limit,
these missing diagrams are replaced  by an expression
which is a splitting function times a tree diagram
with one less leg.
Only one contribution is produced in this limit;
this is because the leg, which in the limit replaces the two legs
of momenta $l_1$ and $l_2$, must always have
negative helicity, otherwise the vertex to which it is attached
would not have the configuration of an MHV vertex.
These missing diagrams can then be rewritten as
a splitting function times a sum of tree MHV diagrams
with $n+1$ external legs. Crucially,
this sum of tree diagrams is precisely such that it combines into a
tree-level $(n+1)$-point amplitude, which is of course
$\eta$-independent.
So we conclude that the sum of the missing diagrams
is actually separately $\eta$-independent,
and hence the one-particle cut diagrams sum to a covariant
expression.

Having settled the one-particle cuts, we move on
to consider two-particle cuts.
These split the one-loop diagram of Figure 1
into two disconnected pieces (see the last diagram in Figure 2).
These are two MHV amplitudes, because the two internal legs
are put on shell by the Feynman cuts. Therefore,
no $\eta$-dependence is produced by these two-particle cut
diagrams.
In conclusion, we have shown that Feynman one-particle and
two-particle cut diagrams are separately covariant. Thus,
by the Feynman Tree Theorem we conclude that the physical
one-loop MHV amplitude is covariant too.

Before moving to more complicated cases,
let us highlight the following points,  anticipating
the general pattern which is emerging.

{\bf 1.} In a one-loop MHV diagram with $v$ vertices
and $n$ external particles
(contributing to an ${\rm N}^{v-2}$MHV amplitude),
the top-cut we can make is necessarily the
$v$-particle cut. This will always be $\eta$-independent
by construction, similarly to the two-particle cut
we have just considered
for the one-loop MHV amplitude.
Notice that this top cut will generically vanish if $v>4$.

{\bf 2.} All $p$-particle cuts which are produced
by the application of the Tree Theorem split
each one-loop MHV diagram into $p$ disconnected pieces
when $p > 1$. In all such cases
we see that amplitudes are produced
on all sides of the cut propagators when the sum over all
MHV diagrams is taken.

{\bf 3.} The case of a one-particle cut is  special since it
generates a connected tree diagram.
Similarly to the case considered before,
one realises that by adding missing diagrams
the one-cut diagrams group into ${\rm N}^{v-1}$MHV amplitudes
with $n+2$ external legs (which are of course covariant).

{\bf 4.}
To see amplitudes appearing on all sides of the cuts,
one has  to sum over all one-loop MHV diagrams.

\begin{figure}[ht]
\label{Figure5}
\begin{center}
\scalebox{0.65}{\includegraphics{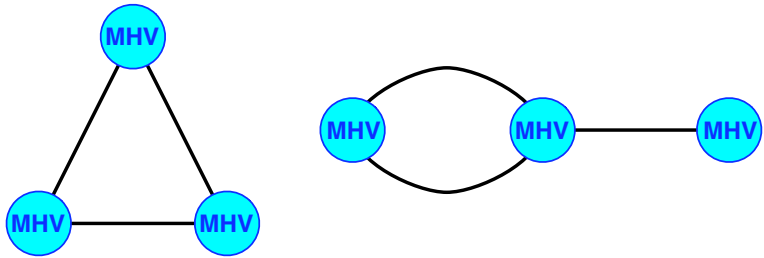}}
\end{center}
\caption{\it These are the MHV diagrams contributing to
a one-loop NMHV scattering amplitude.
}
\end{figure}

\subsection{The NMHV amplitude at one loop}

We now move on and consider the MHV diagrams contributing
to a one-loop NMHV amplitude.
These diagrams are drawn in Figure 4.
The independent one-cut diagrams produced
in the application of the
Feynman Tree theorem are depicted in Figure 5;
the missing diagrams, relevant in the study of one-particle cuts
as explained earlier, are shown in Figure 6.
It is easy to see that
the one-particle cut diagrams of Figure 5 give,
upon inclusion of the missing one-cut diagrams
(drawn in Figure 6),
an ${\rm N}^2$MHV tree amplitude, which is
covariant.
Furthermore,
two-particle cuts, represented in Figure 7,
give an NMHV tree amplitude
joined onto an MHV tree amplitude, and
three-particle cuts give rise to three disconnected
MHV amplitudes, each of which is covariant.

\begin{figure}[ht]
\label{Figure6}
\begin{center}
\scalebox{0.65}{\includegraphics{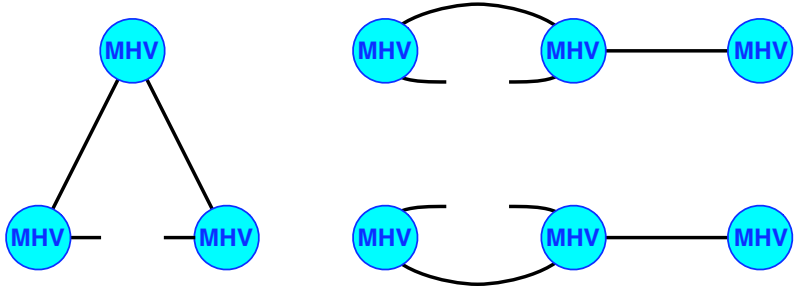}}
\end{center}
\caption{\it This Figure shows the one-particle cut diagrams
generated by cutting open one loop propagator in the diagrams
of Figure 4. Notice that the two diagrams
on the right hand side are independent
and should therefore be included separately.
}
\end{figure}

\begin{figure}[ht]
\label{Figure7}
\begin{center}
\scalebox{0.65}{\includegraphics{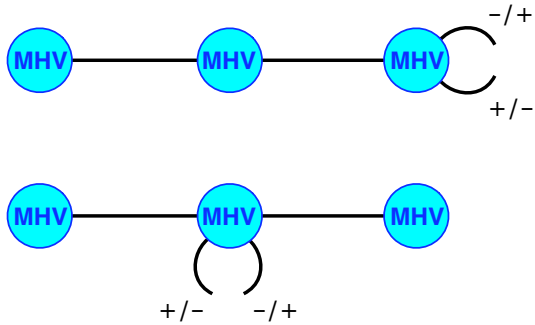}}
\end{center}
\caption{\it Here we draw the missing diagrams corresponding
to a NMHV scattering amplitude at one loop. These are one-particle
diagrams which would never be generated by cutting open one loop propagator
in the one-loop MHV diagrams of Figure 4.
}
\end{figure}

\begin{figure}[ht]
\label{Figure8}
\begin{center}
\scalebox{0.65}{\includegraphics{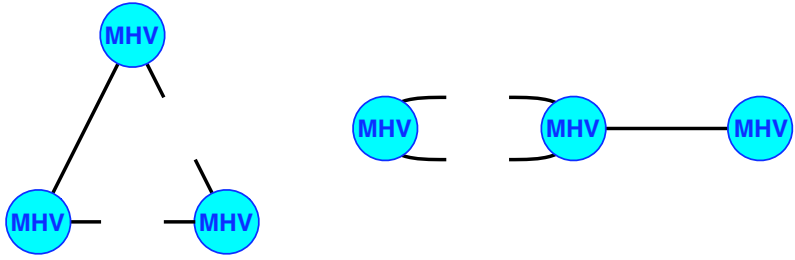}}
\end{center}
\caption{\it
Two-particle cuts of an NMHV amplitude.}
\end{figure}

\subsection{The ${\bf N}^2$MHV amplitude at one loop}

We will conclude this section by considering
the case of the ${\bf N}^2$MHV amplitude at one loop.
This case is general enough to serve as
an illustration for more complicated
cases, which can be treated in a completely similar fashion.
The corresponding MHV diagrams are depicted in Figure 8.
Upon applying the Feynman Tree Theorem, we will produce
Feynman one-, two-, three- and four-particle cut diagrams.
In complete similarity to the cases considered before,
it is easy to see that:
\begin{figure}[ht]
\label{Figure10}
\begin{center}
\scalebox{0.65}{\includegraphics{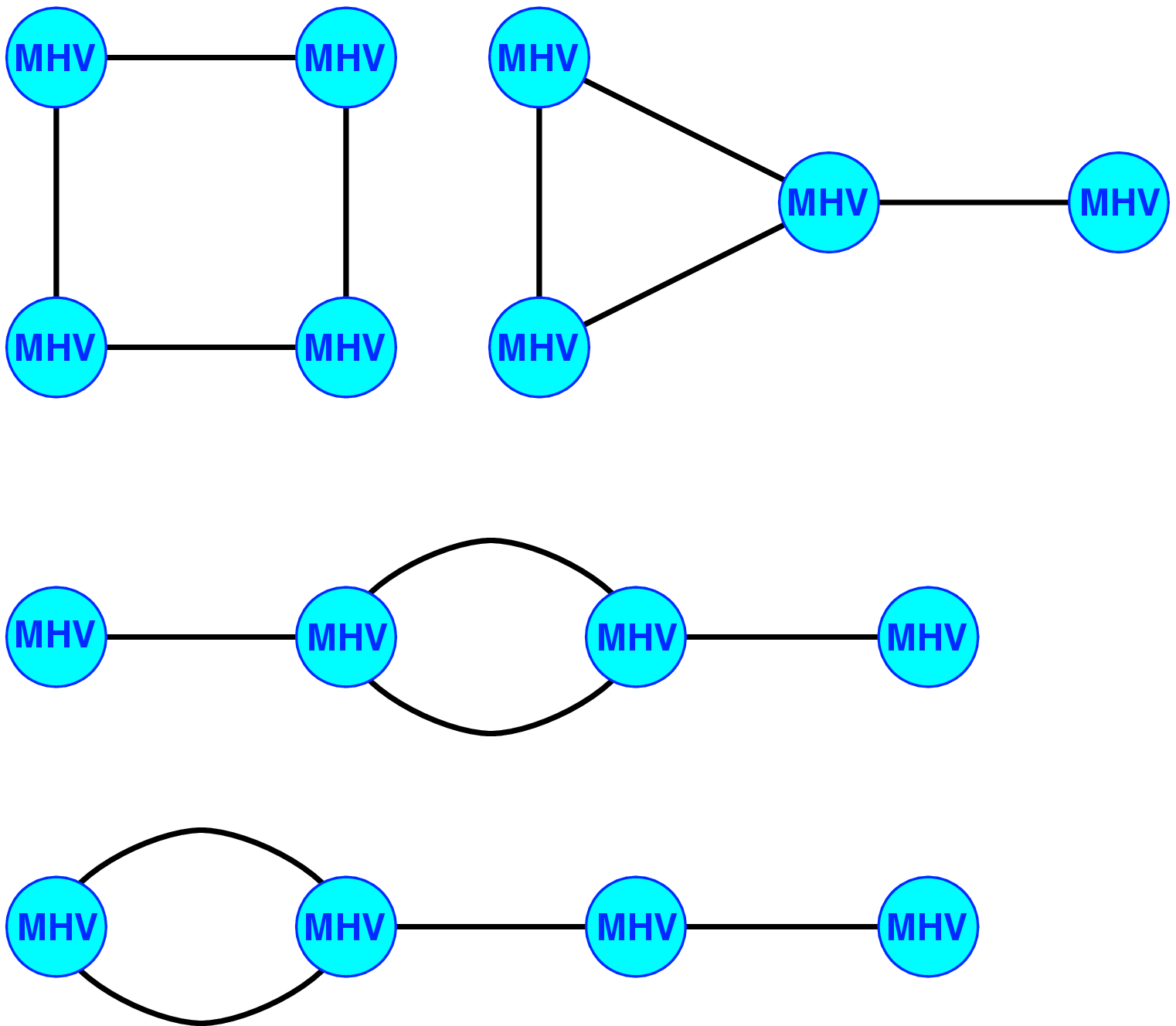}}
\end{center}
\caption{\it MHV diagrams contributing to a
${\rm N}^2$MHV scattering amplitude at one loop.
}
\end{figure}
\begin{itemize}
\item[{\bf 1.}]
The one-particle cut diagrams give, upon inclusion of the
missing one-cut diagrams an ${\rm N}^3$MHV tree amplitude
with $n+2$ external legs, which is covariant;
\item[{\bf 2.}]
Two-particle cuts give rise to two different possibilities:
either an ${\rm N}^2$MHV and an MHV amplitude,
or two NMHV amplitudes;
\item[{\bf 3.}]
Three-particle cuts produce an NMHV amplitude
and two MHV amplitudes;

\item[{\bf 4.}]
Finally, four-particle cuts give rise to four disconnected
MHV amplitudes which are trivially covariant.
\end{itemize}

Again, each set of $p$-particle cut diagrams is covariant.
As a consequence of the Feynman Tree Theorem,
the sum of these sets reproduces the full physical amplitude,
which is therefore covariant.

A final comment before closing this section --
so far we have not been working in any particular
Yang-Mills theory; in particular, we have not required it
to be supersymmetric.
Therefore, the results of this section also show that
the cut-constructible part of amplitudes in pure
Yang-Mills computed using MHV diagrams
is covariant. This is of course in agreement with
the explicit results found in \cite{bbst2} for the
particular case of MHV amplitudes.

%%%%%%%%%%%%%%%%%%%%%%%%%%%%%%%%%%%%%%%%%%%%%%%%%%%%%%%%%%%%%

\section{General Structure of MHV diagrams at one loop}

%%%%%%%%%%%%%%%%%%%%%%%%%%%%%%%%%%%%%%%%%%%%%%%%%%%%%%%%%%%%%

Having shown in the previous sections that
the MHV method at one loop produces covariant expressions, 
we now show that general properties of one-loop MHV diagrams 
are in precise agreement with those arising from 
Feynman diagrams at one loop.

In \cite{bcfw}, Britto, Cachazo, Feng and Witten
gave an elegant proof that MHV diagrams at tree level
reproduce the corresponding amplitude computed from
Feynman diagrams.
Denoting by $\cA_{\rm MHV}^{(n)}$ and
$\cA_{\rm F}^{(n)}$ the results of the
MHV diagram and Feynman diagram calculations,
the authors of \cite{bcfw} notice
that $\cA_{\rm MHV}^{(n)}$ has precisely
the same singularities (simple poles) as
$\cA_{\rm F}^{(n)}$ \cite{csw}.
Of course a single MHV diagram taken in isolation
contains additional unphysical poles coming from the
presence of the reference momentum $\eta$. 
These singularities are of the form
$1/ \lan \l_i \, \l_P \ran$ where
$\l_P^a  := P^{\dot{a} a } \tilde{\eta}_{\dot{a}}$ is the spinor
associated to an off-shell momentum $P$
in the MHV diagram prescription,
and the label $i$ refers to a generic external particle,
with momentum $\l_i \lt_i$.
However, the sum of all tree-level MHV diagrams
is covariant (i.e.~$\eta$-independent),
hence these unphysical poles must cancel out
in the sum $\cA_{\rm MHV}^{(n)}$ \cite{csw,bcfw}.
It then follows that  the difference
$\cA_{\rm F}^{(n)} - \cA_{\rm MHV}^{(n)}$
must be a polynomial in the momenta of the particles.
But scattering amplitudes of $n$ gluons
have dimension $4-n$; so for $n>4$ the polynomial
actually vanishes \cite{bcfw}.
Explicit calculations settle
the issue for the boundary case $n=4$.

The proof of the MHV diagrams method  
at one loop proceeds along similar lines --
with the covariance of the amplitude 
at one loop being a cornerstone of the argument.
Firstly, we observe that by construction
$\cA_{\rm MHV}^{(n)}$ has precisely the same
discontinuities of $\cA_{\rm F}^{(n)}$.
This is manifest in the calculation
of the one-loop MHV amplitude of \cite{bst},
where the two-particle cut of each one-loop
MHV diagram contributing to the amplitude
is directly mapped onto
a (two-particle) unitarity cut of the same
amplitude. For non-MHV  amplitudes, one proceeds
in a way similar to that discussed in previous sections.
Consider a generic non-MHV amplitude computed 
using MHV diagrams at one loop. 
In order to study the discontinuity
across a certain cut, fix the two propagators which
correspond to that particular cut, and cut them.
Then inspect all MHV diagrams contributing to the amplitude
which have that particular two-particle cut. The next step
consists in realising that the sum of these cut-MHV diagrams
is precisely equal to  the two-particle unitarity cut of the
amplitude we want to compute. Hence the discontinuity
across that cut computed using MHV diagrams
equals the physical discontinuity  as computed from
e.g.~Feynman diagrams.
By the same token it follows that 
all generalised cuts are also correct. 

Having shown that the MHV diagrams method gives covariant 
expressions with all the correct cuts (and generalised cuts), 
the last thing to show is that
all the physical poles (collinear and multi-particle)
are also correctly reproduced in an MHV diagram calculation.
Assuming for the moment that this is the case
(we will discuss this in the remainder of this section),
then one could still argue that $\eta$-dependent terms
might give rise to unphysical
discontinuities or poles.
A key point is therefore to prove the
covariance of the amplitude $\cA_{\rm MHV}^{(n)}$ at one loop.
But this is precisely what we have achieved in section 3.
Therefore, no additional, unphysical singularities
can survive in an MHV diagram calculation
at one loop.  If the two functions
$\cA_{\rm MHV}^{(n)}$ and
$\cA_{\rm F}^{(n)}$ have identical discontinuities
and poles, then they must differ at most by polynomial terms.
As in the tree-level proof of \cite{bcfw},
such polynomial terms must vanish on dimensional
grounds for $n>4$.
For $n=4$, where such a polynomial could occur,
the only non-vanishing one-loop
amplitudes in supersymmetric Yang-Mills are the
MHV amplitudes, computed in \cite{bst}.
Hence, one would  have proved that
scattering amplitudes in supersymmetric Yang-Mills
can be equivalently computed
either  with Feynman diagrams or with
MHV diagrams. 
%-- obtaining the same result.

We will discuss collinear and soft limits in the following.
In particular, we will demonstrate that amplitudes calculated 
with MHV diagrams have precisely the expected 
universal collinear factorisation properties.  
We do not have a proof of the correct 
multi-particle factorisation%
\footnote{The naive contributions to the multi-particle 
factorisation formula follow
automatically from the MHV diagrams; 
the difficult piece is to reconstruct
the correct one-loop factorisation functions. 
See e.g.~\cite{Bern:1995ix} for a detailed
discussion of one-loop multi-particle factorisation.} 
at this point; this would be the final piece needed 
to construct a complete proof of equivalence of 
Feynman and MHV diagrams. 
Given the non-trivial checks we will present below, 
we expect this to also follow upon further analysis.

One important comment is in order here.
We should make clear that the following considerations 
will apply to
supersymmetric theories. The reason is that
we will be able to reproduce  expected
physical poles of
the amplitudes only in the presence of supersymmetry.
It is indeed known that for non-supersymmetric Yang-Mills,
a calculation based on MHV diagrams at one loop misses
certain rational terms \cite{csw2}.
In  \cite{bbst2} it was shown for the one-loop MHV amplitude
in pure Yang-Mills that  
the cut-constructible part of the amplitude
is correctly calculated. The arguments presented in this section
indicate that this result extends to the case of
any scattering amplitude in non-supersymmetric Yang-Mills --
so that the cut-constructible part
of non-MHV amplitudes in non-supersymmetric Yang-Mills
can also be reliably computed using MHV diagrams.

\subsection{Proof of
universal collinear factorisation
in the MHV diagrams method}
Now we come back to the issue of reproducing the
expected physical poles.
We will first discuss in detail collinear limits
of generic gluon amplitudes.%
\footnote{Amplitudes with external fermions or 
scalars can be addressed in a completely similar fashion.}

Consider a one-loop scattering amplitude,
$\cA_n^{1-{\rm loop}}$. When the massless legs
$a$ and $b$ become collinear, the amplitude factorises
as \cite{bdk1,Bern:1994cg,Bern:1995ix}
\beqa
\label{coll}
&& \cA_n^{1-{\rm loop}}
(1,\ldots, a^{\l_a}, b^{\l_b}, \ldots, n)
\;
{\buildrel a \parallel b\over
{\relbar\mskip-1mu\joinrel\longrightarrow}}
\\ \nonumber
&& \hspace{1 cm}\sum_{\s} \bigg[
 {\rm Split}^{\rm tree}_{- \s} ( a^{\l_a}, b^{\l_b})
\ \cA_{n-1}^{1-{\rm loop}}
(1,\ldots, (a+ b)^{\s},  \ldots, n)
\nonumber \\
&& \hspace {1.4cm}
\ + \
{\rm Split}^{1-{\rm loop}}_{- \s} ( a^{\l_a}, b^{\l_b})
\ \cA_{n-1}^{\rm tree}
(1,\ldots, (a+ b)^{\s},  \ldots, n)
\bigg]
\ .
\nonumber
\eeqa
${\rm Split}^{\rm tree}$ are the gluon tree-level
splitting functions, whose explicit forms can be found
e.g.~in  \cite{lance}; in particular,
\beqa
 {\rm Split}^{\rm tree}_{-}(a^+, b^+) &=&
{1 \over \sqrt{z(1-z)}} {1 \over \lan a \, b \ran}
\ ,
\\
{\rm Split}^{\rm tree}_{+}(a^-, b^-) &=&
- {1 \over \sqrt{z(1-z)}} {1 \over [ a \, b ]}
\ ,
\eeqa
with $k_a := z k_P$, $k_b := (1-z) k_P$, and
$k_P^2 \to 0$,  in the collinear limit.
${\rm Split}^{1-{\rm loop}}$ is a
supersymmetric one-loop splitting function.
In \cite{ku} and \cite{vittorio}
explicit formulae for this one-loop splitting
function, valid to all orders
in the dimensional regularisation
parameter $\e$, were found.
We quote here the result of \cite{vittorio}:%
\footnote{The result of \cite{ku} is seen to
be identical to that of  \cite{vittorio} after using
equations \eqref{v1} and \eqref{v2}.}
\beq
\label{pippi}
{\rm Split}^{1-{\rm loop}}_{- \s} ( a^{\l_a}, b^{\l_b})
\ = \
{\rm Split}^{\rm tree}_{- \s} ( a^{\l_a}, b^{\l_b})
\ r_1^{[1]} (z)
\ ,
\eeq
where, to all orders in $\e$,
\beq
\label{cl}
r_1^{[1]} (z) \ := \
{c_\Gamma \over \e^2} \Big( {-s_{ab} \over \mu^2} \Big)^{-\e}
\left[  1 \, - \,
\mbox{}_{2}F_1 \left( 1, -\e, 1- \e, {z-1 \over z}\right)
  \, - \, \mbox{}_{2}F_1 \left(1, -\e, 1- \e, {z \over z-1}\right)
\right]
\ ,
\eeq
and
\beq
c_\G \ := \ {\G (1 + \e) \G^2 ( 1 -  \e) \over (4\pi)^{2- \e}
\G(1 - 2 \e)}
\ .
\eeq
Notice that in \eqref{coll} we sum over the
two possible helicities $\s=\pm$.

It is the purpose of
this section to reproduce \eqref{coll},
and in particular the all-orders in $\e$
expressions \eqref{pippi} and \eqref{cl}, previously
derived in \cite{ku,vittorio}.

\subsubsection{Collinear limits in the
MHV diagrams approach at tree level}

Collinear limits of amplitudes in the MHV diagram 
approach at tree level were already studied in the original paper 
\cite{csw}, and found to be in agreement with the known 
results from Feynman diagrams. 
Multi-collinear limits were later analysed in 
\cite{bgkm1,bgkm2}. 
Since the MHV diagram method is not manifestly parity symmetric, 
one needs to distinguish two types of collinear limits \cite{csw}:
\begin{itemize}
\item[{\bf a.}]
Collinear limits where the number of negative helicities
is unchanged; these are the limits
$++ \to +$ and $+- \to -$.
\item[{\bf b.}]
Collinear limits where the number of negative helicities
is reduced by one; these are
$-- \to -$ and $+- \to +$.
\end{itemize}
In both cases, the MHV diagrams contributing to the collinear
limit are only those where the two legs becoming collinear
belong to the same MHV vertex \cite{csw}.
However, in the  cases {\bf a.} and
{\bf b.} the singular behaviour --
encoded in the tree-level splitting functions --
arises from different types of MHV diagrams,
as explained in section 4 of \cite{csw}.
Collinear singularities of type {\bf a.} are directly
inherited from  the MHV vertex to which
the two legs becoming collinear are attached. 
On the other hand,
collinear singularities of type {\bf b.} arise from
special kinds of MHV diagrams, namely those
where the two legs becoming collinear
are attached to a three-point MHV vertex
(connected to another MHV vertex through
a propagator which is going on shell in
the collinear limit).
The underlying reason why the splitting function for
collinear limits of type {\bf b.} arise in this peculiar way
is that amplitudes with only one negative
helicity vanish at tree level.

As we shall see, at loop level the splitting functions
for these two different cases also arise from
different MHV diagrams.
As illustrative examples for the two possibilities,
we will study in detail the two collinear limits
$++ \to +$ (belonging to type {\bf a.})
and $-- \to -$ (belonging to type {\bf b.})
for a generic non-MHV amplitude.
All other cases can be treated in an identical fashion 
and lead to the expected results.

Our diagrammatic analysis will also explain the
universal nature of collinear limits at one loop --
a feature which emerges naturally
in the MHV diagrams approach.
We will be able to derive expressions for the splitting functions 
to all orders in $\e$, matching previous results of 
\cite{ku,vittorio}. This is perhaps surprising -- 
a priori one would expect to find expressions which 
are correct only in the four-dimensional limit, 
that is up to terms which vanish 
when $\e \to 0$.

\subsubsection{``Singular channel'' and ``non-singular
channel'' MHV diagrams}
Before proceeding to analyse the two above-mentioned
cases, we would like to discuss how
the two terms in \eqref{coll} arise in the
MHV diagrams approach. What we are going to see is that,
in some similarity with the unitarity-based derivation
of \cite{david}, these two terms have a clearly separate
diagrammatic origin: the first one arises
from studying MHV diagrams where the kinematical invariant
$s_{ab}:=(k_a + k_b)^2$
(with $s_{ab} \to 0$ in the collinear limit)
corresponds to a ``non-singular'' channel, the second from
diagrams where $s_{ab}$ is a ``singular-channel''.
We define singular channel MHV diagrams to be
those where the following two conditions are
satisfied (see Figure 9):
\begin{itemize}
\item[{\bf 1.}]
the two legs becoming
collinear, $a$ and $b$, belong to a four-point MHV vertex;
\item[{\bf 2.}]
the remaining two legs of this four-point vertex
(which are loop legs) are attached to the same MHV vertex.
\end{itemize}
These diagrams  are quite special,
for the following reason \cite{david}.
By momentum conservation at the four point vertex,
the condition $s_{ab} \to 0$ implies  that
$(L_2 - L_1)^2:= s_{L_1 L_2} \to 0$.
When performing the loop integration, one encounters a region
where the two loop legs $L_1$ and $L_2$ both go on shell,
$L_i \to l_i$, with $l_i^2 =0$ ($i=1,2$).
In this case, the condition  $s_{L_1 L_2} \to 0$
actually implies the collinearity of
$l_1$ and $l_2$. In the singular-channel diagrams,
the two loop legs attached to the same MHV vertex,
and this generates a further collinear singularity in
the MHV vertex  on the right-hand side.%
\footnote{For more details on this point we refer the reader to 
sections 2 and 3 in \cite{david}.}
The conclusion of this discussion is that the collinear limit
of these diagrams
needs to be addressed with special care; importantly,
the collinear limit $s_{ab} \to 0$ must be taken before
the four-dimensional one \cite{david}.
We will see
in the specific calculation performed in the following that it is
precisely diagrams of the singular type which give rise to
the one-loop splitting function (the second term on the 
right hand side of \eqref{coll}).
Furthermore, we also notice that
this particular class of diagrams
gives rise to infrared-divergent contributions,
as explained in \cite{lits}.

Non-singular channel diagrams are those where the two legs
$a$ and $b$ are a proper subset of the legs attached to
a single MHV vertex, or those where $a$ and $b$ belong to a
four-point MHV vertex $\cA_{4, {\rm MHV}} ( a, b, l_2,- l_1)$
but the loop legs $L_1$ and $L_2$ are connected to different
MHV vertices. In this case, even if $L_1$ and $L_2$
become null and collinear, nothing special happens to the
sum of tree MHV diagrams on the right hand side of this
four-point MHV vertex,
precisely because $L_1$ and $L_2$ are not part of
the same MHV vertex.
For the very same reason, diagrams where the two legs
becoming collinear belong to different MHV vertices do not
develop collinear singularities. 
This important fact was noticed
at tree level in \cite{csw},
and the presence of a loop integration does not alter
this conclusion. Therefore, we need not consider
such diagrams in our analysis.

We now move on to derive explicitly the collinear factorisation
$ ++ \to +$ and $ -- \to -$ at one loop.

\subsubsection{$++ \to +$ collinear limit at one loop}

We start our discussion by considering diagrams where
$s_{ab}$ is a non-singular channel.
Let  $\cA_{v, {\rm MHV}}
( \ldots , a^{+}, b^{+}, \ldots, l_2 ,- l_1 ,
\ldots )$ be the MHV vertex to which
the legs $a^+$ and $b^+$
are attached, which by assumption will
be a $v$-point vertex with either $v\geq 5$,
or $v=4$ and the two loop legs are connected to
two different MHV vertices.
When $a^+$ and $b^+$ become collinear, this vertex
has the usual collinear singularity
of a scattering amplitude
\beqa
&& \cA_{v, {\rm MHV}}
( \ldots , a^{+}, b^{+}, \ldots, l_2 ,- l_1 ,
\ldots )
\;
{\buildrel a \parallel b\over
{\relbar\mskip-1mu\joinrel\longrightarrow}}
\nonumber \\ \cr
&&
\hspace{1cm}
{\rm Split}^{\rm tree}_{-} ( a^{+}, b^{+})
\ \cA_{v-1, {\rm MHV}}
( \ldots , (a+b)^{+}, \ldots, l_2 ,- l_1 , \ldots )
\ ,
\eeqa
which is not altered by the fact that $l_1$ and $l_2$
are analytically continued off the mass-shell.
Summing over all MHV diagrams
where $s_{ab}$ is a non-singular channel,
one immediately sees that
a contribution identical to the first term in
\eqref{coll} is generated. This is because
replacing $\cA_{v, {\rm MHV}}
( \ldots , a^{+}, b^{+}, \ldots, l_2 , -l_1 ,
\ldots )$
with a  loop-independent splitting function times
$\cA_{v-1, {\rm MHV}}
( \ldots , (a+b)^{+}, \ldots, l_2 , -l_1 , \ldots )
$
gives that splitting function times
a sum of one-loop MHV diagrams which precisely
add up to $\cA_{n-1}^{1-{\rm loop}}
(1,\ldots, (a+ b)^{+},  \ldots, n)
$.

Next we  move to the singular-channel diagrams
(such as the one depicted in Figure 9),
i.e.~diagrams where  the legs $a$ and $b$  belong to a
four-point MHV vertex, and the two remaining loop legs are
attached to the same MHV vertex.

\begin{figure}[ht]
\label{Figure9}
\begin{center}
\scalebox{0.65}{\includegraphics{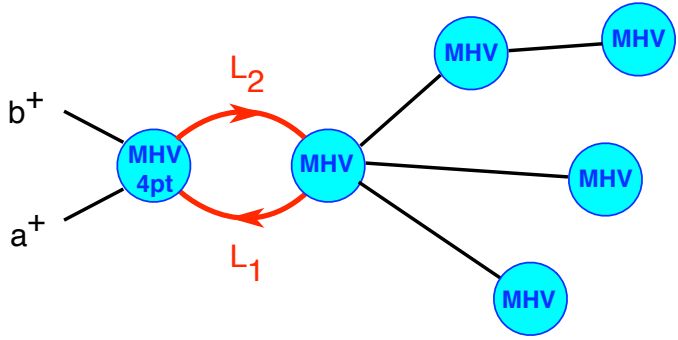}}
\end{center}
\caption{\it
A schematic example of a one-loop
MHV diagram contributing to a generic
non-MHV one-loop amplitude where
$s_{ab}$ is a singular channel. In the collinear
limit $a \parallel b$, diagrams of this type
generate the second term on the right hand side of  
\eqref{coll}.
}
\end{figure}

The diagram represented in Figure 9 is one of the one-loop MHV
diagrams calculated in
\cite{bst}. The fact that one or more of the legs attached to the
MHV vertex on the right hand
side attach to further vertices
does not actually alter the calculation of \cite{bst}. Because
$a$ and $b$ have positive helicity, only gluons can run in the loop.
The loop integrand
involves the function \cite{bdk1,bst}
\beq
\label{funct-R}
\hat{\cR} \ := \ {
\lan a - 1 \, a \ran \lan l_2 \, l_1\ran \over \lan a-1 \,
l_1 \ran \lan -l_1 \, a \ran}\ { \lan b  \, b + 1 \ran \lan
l_1 \, l_2 \ran \over \lan b  \, l_2 \ran \lan -l_2  \, b + 1
\ran} \ ,
 \eeq
where the spinors $l_1$ and $l_2$ are those associated
to the off-shell loop momenta $L_1$ and $L_2$
according to the prescription of \cite{csw},%
\footnote{In section 5.2 we briefly review this
off-shell prescription, and also present
two derivations of the one-loop integration measure of \cite{bst}
to be used in calculating one-loop MHV diagrams such as the one in
Figure 9. We refer the reader to that section, and
to sections 3--5 of \cite{bst} for more details.}
that is
\beq
L_{i; \a, \da} \ =
\ l_{i \a} \llt_{i \da}  \, + \,
z_i \, \eta_{\a}\tilde{\eta}_{\da} \ ,
\qquad i=1,2
\ .
\eeq
Notice that momentum conservation
requires $L_2 - L_1 + P_L =0$, where
in the singular channel $P_L = k_a + k_b$.

Using the Schouten identity,  we can recast
$\hat{\cR}$ as
\beq \label{sum-R}
 \hat{\cR} \ = \
-  \cR( a, b ) \, -  \, \cR( a -1, b +1) \, + \,
\cR( a , b + 1) \, + \,  \cR( a-1 , b)
\ ,
 \eeq
where $\cR( i, j )$ is  the homogeneous function of
the spinors $l_1$ and $l_2$ given by
\beq
\cR (i\, j) \, := \,
{\lan i \, l_2 \ran \over \lan i \, l_1
\ran} \, {\lan j \, l_1 \ran \over \lan j \, l_2 \ran}
\ .
\eeq
We now rewrite  the integrand in terms of the scalar functions
appearing in the bubble, triangle and box integrals.
Firstly, we notice that%
\footnote{In the following formula we will omit
a term proportional to an $\epsilon$-tensor contracted with
four momenta, which
vanishes upon integration.}
\beqa\label{Mai}
\cR (i\, j) \, &= & {\lan i \, l_2 \ran \,
[l_2 \, j] \, \lan j \, l_1 \ran \, [l_1 \, i]\over \lan i \, l_1 \ran
\, [l_1 \, i] \,  \lan j \, l_2 \ran\, [l_2 \, j]} \ = \ { {\rm
Tr} \left[ {1\over 2}(1 - \gamma^5) \,  \hat{i} \,
\hat{l}_2 \, \hat{j} \, \hat{l}_1 \right]
\over 4  (l_1 i) (l_2 j)}
\\ [8pt]\nonumber
& =& {   (l_1 \, i) (l_2 \, j) + (l_1 \, j) (l_2 \, i) -
(l_1 \, l_2) (i\, j)  \over 2 (l_1  i) (l_2  j)} \ .
\eeqa
Momentum conservation can be re-written in terms
of $l_1$ and $l_2$ as \cite{bst}
\beq
\label{abc}
l_2 - l_1 + P_{L;z}\ = \ 0
\ ,
\eeq
with
$P_{L;z} := P_L - z \eta$, and  $z := z_1 - z_2$.
Using \eqref{abc} we can rewrite
\beq\label{Mai2}
 \cR (i\, j) \ = \ 1 \, + \, {1\over 2}
\bigg[ - {(i P_{L;z}) \over (i l_1) }\,  + \,
 {(j P_{L;z}) \over (j l_2) } \bigg] \, + \,
{1\over 4} { P_{L;z}^2 (ij) \, - \, 2 (i P_{L;z})
(j P_{L;z}) \over (il_1) (jl_2)}
\ .
\eeq
Upon summing over the four terms in
\eqref{sum-R},
bubbles (corresponding to the first term on the right-hand side in
\eqref{Mai2}) and triangle integrals (the term in square brackets
in \eqref{Mai2}) cancel, so we can replace
the function $\cR$ by an
effective  function $\cR^{\rm eff}(i,j)$ containing only
the contributions of the box functions,
\beq
\label{Reff}
\cR^{\rm eff}(i,j;z) \ := \
{1\over 4} { P_{L;z}^2 (ij) \, - \, 2 (i P_{L;z})
(j P_{L;z}) \over (il_1) (jl_2)}
\ .
\eeq
Specifically, these turn out to be two-mass easy box functions,
that is box functions with two non-adjacent massless legs, $i$
and $j$ in \eqref{Reff}, with the remaining two legs being
massive.
Furthermore, due to the identity
\beqa
\label{identity}
4(Pi)(Pj) - 2P^2(ij) &= &(P+i)^2(P+j)^2 \, - \, P^2(P+i+j)^2
\\ \nonumber
&:=& st \, - \, P^2 Q^2
 \ ,
\eeqa
valid for any momentum $P$,
the function \eqref{Reff} is invariant under
\beq P_{L;z}
\to P_{L;z}  + \a i  + \b j \ ,
\eeq where $\a$ and $\b$ are arbitrary
numbers.
Thus, choosing $\eta$ to be either $i$ or $j$
one can replace $\cR^{\rm eff}(i,j;z)$ by
$\cR^{\rm eff}(i,j;0)$.

In \cite{bst} it was shown that each  of the four terms in
\eqref{sum-R} (where, at this point, $\cR$ can be replaced
by $\cR^{\rm eff})$ gives rise to a dispersion integral 
in the $s_{ab}$ channel of the $s_{ab}$-cut of four different
box functions -- specifically, $\cR (i,j)$ correspond to 
a two-mass easy box with massless legs $i$ and $j$.
We would like to point out that 
$\eta$ can be chosen separately for each box function,
so as to remove the $\eta$-dependence in the
$\cR^{\rm eff}$ function. 
Indeed in \cite{bst} it was shown (numerically) 
that $\eta$-dependence in the dispersion integral
in one of the four channels of a certain box function
cancels when combined with the integrals for the other
three channels, for the terms which are 
singular and finite in $\e$.% 
\footnote{So in fact at the final stage of performing 
all the dispersion integrals $\eta$ can effectively 
be set to zero. 
Note, however, that for all previous stages 
of the calculation $\eta$ has to be kept non-zero.}
We have now extended this result by 
further verifying $\eta$-independence 
for the all orders in $\e$ expression 
of the box function.

\begin{figure} [ht]
\label{fig2}
\vspace{.2in}
\centerline {
\includegraphics[width=3in]{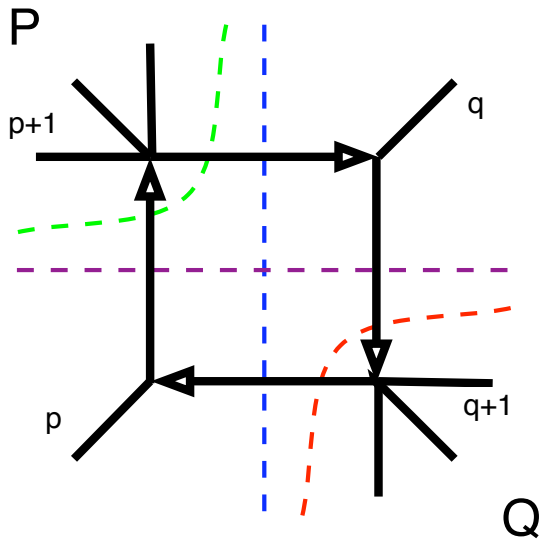}
} \vspace{.2in} \caption{\it The two-mass easy box function
$F^{\rm 2me}$, whose all-order in $\e$ expression is
given in \eqref{box-allorders} and further
studied in the Appendix.
The vertical (horizontal) cuts correspond to the
$s$-channel ($t$-channel) cuts,
and the upper left (lower right) corner cuts to the
$P^2$-channel ($Q^2$-channel) cuts.}
\end{figure}

The four terms in \eqref{sum-R} give rise to four
different boxes in the $s_{ab}$-cut, which are then
integrated with the appropriate dispersive measure
in the $s_{ab}$-channel
(see section 5.2 and \cite{bst} for further details).
Thus, what we are seeing here is that the
one-loop  collinear singularity
arises precisely from summing 
these four dispersion integrals. 

A two-mass easy box is uniquely identified
by specifying the two massless legs
(see Figure 10 for the definition of the various
kinematical invariants).
For the four boxes in \eqref{sum-R},
the massless legs are
{\bf I.} $a$ and $b$, {\bf II.} $a-1$ and $b+1$,
{\bf III.} $a$ and $b+1$, {\bf IV.} $a-1$ and $b$.
Now we make use of the following all-order in $\e$
expression for the two-mass easy box function,
also discussed in the Appendix,%
\footnote{The  four-dimensional limit of 
\eqref{box-allorders}, written explicitly in 
\eqref{niceonecyril}, was derived in \cite{bst} 
as the sum of four dispersion integrals 
(one for each cut of the box function) 
of appropriate phase space integrals.
By keeping these phase space integrals 
to all orders in $\e$, one arrives at \eqref{box-allorders}.
As mentioned before, we have also performed extensive 
numerical checks that $\eta$ dependence cancels between the four
dispersion integrals even  if we work with the 
all-order expressions of the phase space integrals.
See the Appendix for further details on
and alternative forms of the all-order 
in $\e$ two-mass easy box functions, 
and section 5 of \cite{bst} for a discussion 
of the analytic continuation to the physical 
region of this expression.
}
\beqa
\nonumber
&&\hspace{-0.7cm}
\label{box-allorders}
 F^{\rm 2me} (s, t, P^2, Q^2) \,= \,
-{ c_{\G} \over \e^2}
\left[
 \Big( {-s \over \mu^2} \Big)^{-\e} \,
\mbox{}_{2}F_1 \left( 1, -\e, 1- \e, as \right)
\, + \,
\Big( {-t \over \mu^2} \Big)^{-\e}
\mbox{}_{2}F_1 \left( 1, -\e, 1- \e, at \right)
\right.
\\  \cr
&&  \qquad - \,
\left.\Big( {-P^2 \over \mu^2} \Big)^{-\e}
\,
\mbox{}_{2}F_1 \left( 1, -\e, 1- \e, aP^2 \right)
\, - \,
 \Big( {-Q^2 \over \mu^2} \Big)^{-\e} \,
\mbox{}_{2}F_1 \left( 1, -\e, 1- \e, a Q^2 \right)
\right]
\, .
\eeqa
Here
\beq
\label{a-deff}
a \ := \ {P^2 + Q^2 - s - t \over P^2 Q^2 - st} \ = \
{2 (p q) \over P^2 Q^2 - st}
\ ,
\eeq
$p$ and $q$ are the two massless legs, and 
we have defined 
\beq
s\ := \ (P+ p)^2 \ , \qquad \quad t\ := \ (P+q)^2
\ . 
\eeq
The first contribution vanishes in the collinear limit,
whereas the other three give a contribution
\beqa
\label{hypersum}
&& { c_{\G} \over \e^2} \Big( {-s_{ab} \over \mu^2} \Big)^{-\e}
\Big[ - \mbox{}_{2}F_1
\left( 1, -\e, 1- \e, a_{\rm II} s_{ab} \right)
\, + \,
\mbox{}_{2}F_1 \left( 1, -\e, 1- \e, a_{\rm III} s_{ab} \right)
\nonumber  \\ [4pt]
&&
\qquad \qquad\qquad
\  + \,
\mbox{}_{2}F_1 \left( 1, -\e, 1- \e, a_{\rm IV} s_{ab} \right)
\Big]
\ ,
\eeqa
multiplied by a prefactor
which is easily seen to be
\beq
{\rm Split}^{{\rm tree}}_{- \s} ( a^{\l_a}, b^{\l_b})
\ \cA_{n-1}^{\rm tree}
(1,\ldots, (a+ b)^{\s},  \ldots, n)
\ .
\eeq
In \eqref{hypersum}
$a_{\rm I}, \ldots, a_{\rm IV}$ are the
expressions for the parameter
$a$ defined in \eqref{a-deff}
appropriate for each of the
four boxes.
With $k_a := z (k_a + k_b)$, $k_b := (1-z) (k_a + k_b)$
and $(k_a + k_b)^2:= s_{ab} \to 0$ in the collinear limit,
we find  that
\beqa
a_{\rm II} \, s_{ab}
& {\buildrel a \parallel b\over
{\relbar\mskip-1mu\joinrel\longrightarrow}}
&
0
\ ,
\nonumber\\ [6pt]
a_{\rm III} \, s_{ab}
& {\buildrel a \parallel b\over
{\relbar\mskip-1mu\joinrel\longrightarrow}}
&{z \over z-1}
\ ,
\nonumber \\ [6pt]
a_{\rm IV} \, s_{ab}
& {\buildrel a \parallel b\over
{\relbar\mskip-1mu\joinrel\longrightarrow}}
& {z-1 \over z}
\ .
\eeqa
Using
$\mbox{}_{2}F_1 \left( 1, -\e, 1- \e, 0 \right) = 1$,
we can immediately re-write
\eqref{hypersum} as
\beq
{c_{\G} \over \e^2} \Big( {-s_{ab} \over \mu^2} \Big)^{-\e}
\left[
-1 \, + \,
\mbox{}_{2}F_1 \left( 1, -\e, 1- \e, {z-1 \over z}\right)
  \, + \, \mbox{}_{2}F_1 \left(1, -\e, 1- \e, {z \over z-1}\right)
\right]
\ ,
\eeq
thus reproducing the
function $r_1^{[1]} (z)$
defined in \eqref{cl}, and hence the
one-loop splitting function to all orders in
$\e$,
for the case of the collinear limit
$++ \to +$.

%%%%%%%%%%%%%%%%%%%%%%%%%%%%%%%%%%%%%%%%%%%%%%%%%%%%%%%%%%%%

\subsubsection{$-- \to -$ collinear limit at one loop}
As in the previous case, we can neatly separate
the diagrams contributing to the tree-level splitting functions
from those contributing to the one-loop splitting functions,
first and second terms in \eqref{coll}, respectively.
We begin by considering the diagrams contributing to
the term in \eqref{coll} containing the tree-level
splitting function (see Figure 11).
These are generated by those MHV diagrams where
the legs becoming collinear, $a^-$ and $b^-$,
are joined to a three-point MHV vertex.
This MHV vertex is then connected to the rest of the diagram.
It is immediate to realise that,
upon taking the collinear limit,
these diagrams give a contribution
${\rm Split}^{\rm tree}_{+} ( a^{-}, b^{-})
\ \cA_{n-1}^{1-{\rm loop}}
(1,\ldots, (a+ b)^{-},  \ldots, n)$.
\begin{figure}[ht]
\begin{center}
\scalebox{0.65}{\includegraphics{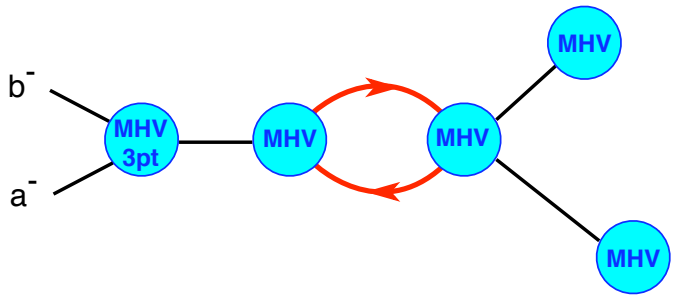}}
\end{center}
\caption{\it
A schematic  example of a one-loop
MHV diagram contributing to
the non-singular channel diagrams
for the $-- \to -$ collinear limit,
giving rise to the contribution
 ${\rm Split}^{\rm tree}_{+} ( a^{-}, b^{-})
\ \cA_{n-1}^{1-{\rm loop}}
(1,\ldots, (a+ b)^{-},  \ldots, n)$.
}
\end{figure}

We now move on to consider the diagrams which are
at the origin of the term
${\rm Split}^{1-{\rm loop}}_{+} ( a^{-}, b^{-})
\ \cA_{n-1}^{\rm tree}
(1,\ldots, (a+ b)^{-},  \ldots, n)$.
These diagrams, represented in Figure 12, are those where
the negative-helicity gluons $a^-$ and $b^-$
are attached to a four-point MHV vertex. Furthermore,
the remaining two legs of this vertex are connected
to a three-point MHV vertex to form a loop.
The remaining leg of this three-point MHV vertex is then attached
to the rest of the diagram.
The need for a three-point vertex is due to the following fact.
As the collinear limit is taken, the MHV vertex on the
left hand side produces a contribution $\lan a \, b \ran^3$
which vanishes in the limit and would make the one-loop diagram
vanish as well. The exception is when the second vertex is
a three-point one; this is because the propagator attached to
this vertex is in this case $i/(k_a+k_b)^2$.
It is also clear that this mechanism is actually similar
to that generating the tree-level splitting function with the
same helicity configuration, hence
${\rm Split}^{1-{\rm loop}}_{+} ( a^{-}, b^{-}) \propto
{\rm Split}^{\rm tree}_{+} ( a^{-}, b^{-})$.
Furthermore, by summing over all possible
MHV diagrams which include such
a sub-diagram we also produce a term
$\cA_{n-1}^{\rm tree} (1,\ldots, (a+ b)^{-},  \ldots, n)$.

\begin{figure}[ht]
\begin{center}
\scalebox{0.65}{\includegraphics{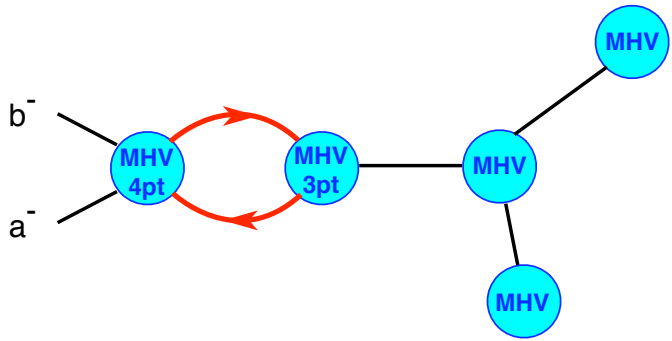}}
\end{center}
\caption{\it
A schematic example of a one-loop
MHV diagram contributing to
the singular channel diagrams
for the $-- \to -$ collinear limit,
giving rise to the term
 ${\rm Split}^{1-{\rm loop}}_{+} ( a^{-}, b^{-})
\ \cA_{n-1}^{\rm tree}
(1,\ldots, (a+ b)^{-},  \ldots, n)$.
}
\end{figure}

The calculation of the diagram in question proceeds along
lines very similar to that followed
for the $++ \to +$ collinear limit, so we will not
spell it out. We would only like to point out one
aspect of it which we believe
is worth mentioning.  In performing the algebra,
one encounters ratios such as
$(k_a k)/ (k_b  k)$, where $k$
is the null momentum associated to the internal
leg attached to the three-point vertex, which carries
momentum $k_a + k_b$. $k$ is defined by the
usual off-shell prescription
$k_a + k_b = k + z_{ab} \eta$, from which 
\beq
k \ = \ k_a + k_b \, - \, {(k_a k_b)  \over 
(k_a \eta)  + (k_b  \eta)}\eta
\ . 
\eeq
It is then easy to see that 
\beq
{k_a k\over  k_b  k} \ = \ 
{k_b \eta \over k_a \eta} 
\ , 
\eeq
so that, in the collinear limit, such ratios give
\beq
\label{clout}
{k_a k\over  k_b  k}
\
{\buildrel a \parallel b\over
{\relbar\mskip-1mu\joinrel\longrightarrow}}
\
{(1-z)  \, (k_P \eta) \over z (k_P \eta)} \ = \
{1- z  \over z }
\ ,
\eeq
where $k_P^2 = (k_a + k_b)^2 \to 0$ and 
$k_a = z k_P$, $k_b = (1-z) k_P$  
in the collinear limit.
Notice that $k$ is null just because of the
CSW prescription -- that is independently
of any collinear limit.

%%%%%%%%%%%%%%%%%%%%%%%%%%%%%%%%%%%%%%%%%%%%%%%%%%%%%%%%%%%%
\subsection{Soft limits of amplitudes with MHV diagrams}

We now turn to discuss soft limits of amplitudes.

\subsubsection{Tree level}
Limits of amplitudes when the momentum 
of one of the particles goes to zero, i.e.~soft limits, 
have a universal behaviour which is captured by soft 
functions (see e.g.~\cite{lance}). 
At tree level, one has
\beq
\label{soft-tree}
\cA_n^{\rm tree}
(1,\ldots, a, s, b, \ldots, n)
\ 
{\buildrel k_s \to 0 \over
{\relbar\mskip-1mu\joinrel\longrightarrow}}
\ 
{\rm Soft}^{\rm tree} ( a, s, b)
\ \cA_{n-1}^{\rm tree}
(1,\ldots, a , b, \ldots, n)
\ , 
\eeq
where ${\rm Soft}^{\rm tree} ( a, s, b)$ is a 
tree-level soft (or eikonal) function, 
\beq
\label{soft-funct-tree}
{\rm Soft}^{\rm tree} ( a, s^+, b) \ = \ 
{\lan a \, b \ran \over 
\lan a \, s \ran \, \lan s \, b \ran }
\ , \qquad 
{\rm Soft}^{\rm tree} ( a, s^-, b) \ = \ 
- \,
{ [ a \, b ] \over 
[ a \, s ]  \, [ s \, b ]  }
\ . 
\eeq
It is easy to see that, at tree level, the 
MHV diagram method precisely reproduces the 
expected behaviour 
\eqref{soft-tree} for the amplitudes, with 
the soft functions given in \eqref{soft-funct-tree}. 
In similarity with the collinear limits, particular care 
is required when the soft gluon 
has negative helicity, and therefore
the number of negative helicities is reduced by one.
In this case, two diagrams are  
relevant for the process. The first one has an 
MHV three-point vertex with external particles $a$ and $s$
($s$ is the leg whose momentum is becoming soft), 
connected to an MHV vertex 
to which the leg $b$ is attached so as to preserve the cyclic 
ordering $a$, $s$, $b$. The second diagram has 
$s$ and $b$ attached to a three-point MHV vertex, 
connected to a second MHV vertex to which $a$ belongs. 
Summing up these two diagrams one obtains, in the soft limit, 
the second expression in \eqref{soft-funct-tree}. 

\begin{figure}[ht]
\begin{center}
\scalebox{0.65}{\includegraphics{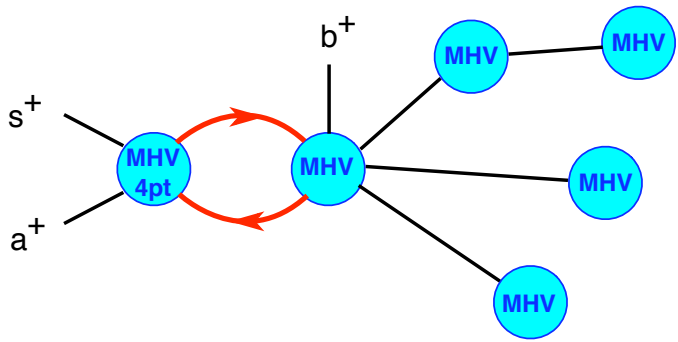}}
\end{center}
\caption{\it
A schematic  example of a one-loop
MHV diagram contributing to
the one-loop soft function 
 ${\rm Soft}^{1-{\rm loop}} ( a^{+},s^+,  b^{+})$.
}
\end{figure}

\subsubsection{One loop}
The behaviour of one-loop scattering amplitudes 
when one of the legs becomes soft is quite similar to 
the collinear behaviour (see \eqref{coll}), 
\beqa
\label{soft-1loop}
&& \cA_n^{1-{\rm loop}}
(1,\ldots, a, s, b, \ldots, n)
\;
{\buildrel k_s \to 0 \over
{\relbar\mskip-1mu\joinrel\longrightarrow}}
\\ \nonumber \cr 
&& 
\hspace{2.3cm}
 {\rm Soft}^{\rm tree} ( a, s, b)
\ \cA_{n-1}^{1-{\rm loop}}
(1,\ldots, a ,  b, \ldots, n)
\nonumber \\ \cr
&& 
\hspace {1.7cm}
\ + \
{\rm Soft}^{1-{\rm loop}} ( a, s, b )
\ \cA_{n-1}^{\rm tree}
(1,\ldots, a,  b \ldots, n)
\ , 
\nonumber
\eeqa
where ${\rm Soft}^{1-{\rm loop}} ( a, s, b )$ 
is a one-loop soft function 
\cite{Kunszt:1993sd,Bern:1991aq,ES,Bern:1993mq,KST,Bern:1994fz}.
We will now obtain the all orders in $\e$ expression 
for the one-loop soft function found in \cite{vittorio2,vittorio}, 
in the case where the helicities of 
the gluons $a$, $s$ and $b$ are all positive.

\begin{figure}[ht]
\begin{center}
\scalebox{0.65}{\includegraphics{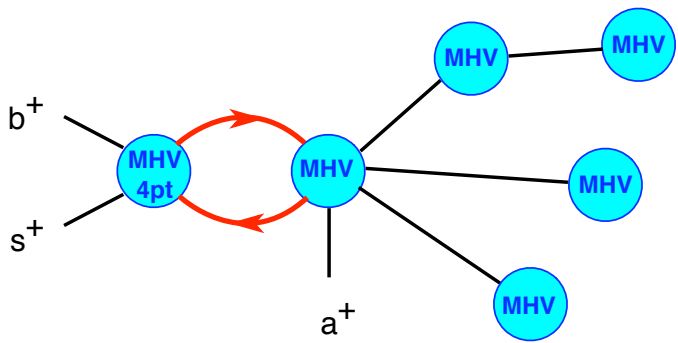}}
\end{center}
\caption{\it
The second class of MHV diagrams 
contributing to 
${\rm Soft}^{1-{\rm loop}} ( a^{+},s^+,  b^{+})$.
}
\end{figure}

The relevant one-loop MHV diagrams are drawn 
in Figures 13-15.
\begin{figure}[ht]
\begin{center}
\scalebox{0.65}{\includegraphics{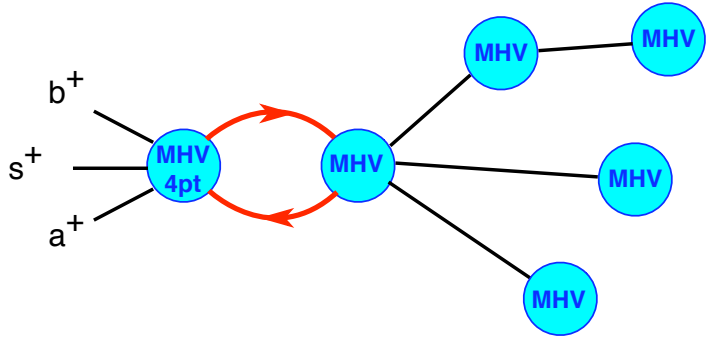}}
\end{center}
\caption{\it
The third class of one-loop MHV diagrams 
contributing to 
${\rm Soft}^{1-{\rm loop}} ( a^{+},s^+,  b^{+})$.
}
\end{figure}
After performing the spinor algebra as in section 4.1, 
one can see that the term singular in the soft limit
gives
\beq
{\rm Soft}^{1-{\rm loop}} (a, s, b)
\ = \ 
 {\rm Soft}^{\rm tree} (a, s, b)
\, 
\left( B_{\rm I} \, + \, B_{\rm II} \, + \, B_{\rm III}
\right) 
\ , 
\eeq
where $B_{\rm I}$, $B_{\rm II}$ and  $B_{\rm III}$ 
are three dispersion integrals in the $s$, $t$ and $P^2$ cut 
of a  one-mass box function with adjacent massless legs 
$a$, $s$ and $b$. For this box, the parameter $a$ defined in 
\eqref{a-deff} is equal to $a = - s_{ab} / (s_{as} s_{sb})$, 
with $s_{ij} := (k_i + k_j)^2$. 
Using the form \eqref{444} for the all-orders box function, 
where $- as_{ab} /(1-as_{ab}) \to 1$ in the soft limit,  
along with 
\beq
\mbox{}_{2}F_1 \left( -\e, -\e, 1- \e, 1 \right)
\ = \ {\pi \e \over \sin ( \pi \e)} 
\ , 
\eeq
we get 
\beq
{\rm Soft}^{1-{\rm loop}} (a, s, b)
\ = \ 
 {\rm Soft}^{\rm tree} (a, s, b)
\, 
\left( -{c_\G \over \e^2} {\pi \e \over \sin ( \pi \e)} 
\right) \, \Big(-{ s_{ab}  \over  s_{as} s_{sb}}\m^2\Big)^\e
\ . 
\eeq
This is in agreement with the results of 
\cite{vittorio2,vittorio}.

\section{Further applications of the Feynman Tree Theorem}

We have seen in previous sections that the
Feynman Tree Theorem \eqref{ftt} leads us to consider
various multiple cuts of a loop diagram.
As reviewed in section 2, these cuts assign a specific
sign to the energy of the cut line; using the
``retarded" version of the theorem,
the sign of the time components of all cut lines is negative,
as prescribed by the second term in  \eqref{decret}.
By summing over these multiple cuts one obtains the
physical amplitude  -- this is the content
of the Feynman Tree Theorem, as expressed in
\eqref{ftt}.

In this section we would like to present
concrete examples of the application of the Feynman Tree Theorem.
The first one is the calculation of a bubble diagram
in a generic scalar theory, the second is the
calculation of the one-loop  MHV scattering amplitudes
of \cite{bst}.

\subsection{Cutting bubbles with the Feynman Tree Theorem}

Here we present a simple application of the
Feynman Tree Theorem to the calculation of
a bubble diagram. We will first evaluate it
in a standard way, and then by applying
the Feynman Tree Theorem.

\begin{figure}[ht]
\label{Figure1}
\begin{center}
\scalebox{0.65}{\includegraphics{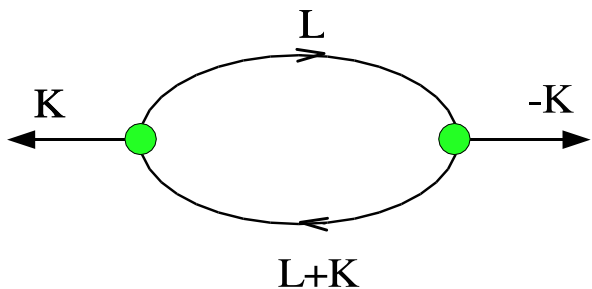}}
\end{center}
\caption{\it
A bubble diagram, which we firstly evaluate directly,
and then by making use of the Feynman Tree Theorem.
The two results are in perfect agreement.
}
\end{figure}

The bubble diagram we consider is represented in Figure 16,
and is given by
\beq
\label{bubble}
\cB_F (K^2) \ := \
\int\!{d^{D}L\over (2\pi)^{D}} \ \D_F(L) \, \D_F (L+ K)
\,
f (L, K)
\ ,
\eeq
where $\D_F$ are  Feynman propagators, given in \eqref{ofey2}.
We can think of \eqref{bubble} as a Feynman diagram
in some scalar field theory; the function $f (L, K)$
of the loop momentum $L$ and
external momentum $K$
in the numerator of \eqref{bubble} is
generated by the interaction vertices.
The presence of this function is however
irrelevant for the following discussion,
thus we will set $f = 1$.
We perform the integration in $D=4-2\e$ dimensions
to regulate divergences in the integral \eqref{bubble};
for $\e\neq 0$ the integration is convergent.
Finally, notice that Lorentz covariance requires $\cB_F$
to be a function of $K^2$.

\subsubsection{Standard calculation of the bubble}

We begin by  first performing
the $L_0$ integration in \eqref{bubble}.
The Feynman propagators can be rewritten using \eqref{ofey}.
Therefore we see immediately that the integrand of
\eqref{bubble} has four simple poles in the complex $L_0$ plane,
at the following locations:
\beqa
&{\bf a. }\ \
L_0 = \o_L - i \vare
\ ,
\qquad \quad
&{\bf b.}\ \
L_0 = - \o_L + i \vare \ ,
\\ \nonumber \cr
& \ \ \ \ \ \ \
{\bf c. }\ \
L_0 = - K_0 - \o_{L+K} + i \vare
\ , \quad
&{\bf d.}\ \
L_0 = - K_0+ \o_{L+K} - i \vare
\ ,
\eeqa
where $\o_L := \sqrt{|\vec{L}|^2 + M^2}$.

We see that the two poles {\bf a.} and {\bf d.}
lie in the lower half $L_0$ plane, whereas
the remaining two are in the upper half $L_0$ plane.
We can close the integration contour
with either a large semicircle above the real axis, 
or one below the real axis,
each time enclosing two of the four poles.
Choosing to close the contour below, we get
\beq
\label{af}
\cB_F \ = \ -2\pi i \ \Big(
{\rm Res}_{\bf a} \, + \, {\rm Res}_{\bf d}
\Big)
\ , 
\eeq
where
\beqa
\label{resa}
{\rm Res}_{\bf a} &=&
\int\!{d^{D-1}\vec{L} \over (2\pi)^D}
\ \,
\lim_{ L_0 \to \o_L - i \vare}
\,
(L_0 - \o_L + i \vare) \, \D_F (L) \, \D_F (L+K)
\\ \nonumber \cr
&=&
{1\over 2\pi}
\int\!{d^{D-1}\vec{L} \over (2\pi)^{D-1} }
\
{i\over 2 \o_L}{i\over 2 \o_{L+K}}
\
\bigg[
{1\over K_0 + \o_L - \o_{L+K}} \, - \,
{1\over K_0 + \o_L  + \o_{L+K} - i \vare}
\bigg]
\ ,
\\ \cr \cr
\label{resd}
{\rm Res}_{\bf d} &=&
\int\!{d^{D-1}\vec{L} \over (2\pi)^D}
\ \,
\lim_{ L_0 \to -K_0 + \o_{L+K} - i \vare}
\,
(L_0 + K_0 - \o_{L+K} + i \vare ) \, \D_F (L) \, \D_F (L+K)
\\ \nonumber \cr
&=&
{1\over 2\pi}
\int\!{d^{D-1}\vec{L} \over (2\pi)^{D-1}}
\
{i\over 2 \o_L}{i\over 2 \o_{L+K}}
\
\bigg[
{1\over -K_0 + \o_{L+K}  - \o_{L}}\,  - \,
{1\over -K_0 + \o_{L+K} + \o_{L} - i \vare}
\bigg]
\ .
\eeqa
Substituting \eqref{resa} and \eqref{resd} into
\eqref{af} we get
\beq
\label{bubfin}
\cA_F \ = \ {i\over 4}
\int\!{d^{D-1}\vec{L} \over (2\pi)^{D-1} }
\ {1\over  \o_L \, \o_{L+K}}
\,
\bigg[
{1\over -K_0 - \o_{L}  - \o_{L+K} + i \vare}\,  - \,
{1\over -K_0 + \o_{L} + \o_{L+K} - i \vare}
\bigg]
\ ,
\eeq
which is our final result.%
\footnote{For our purposes \eqref{bubfin} will
be sufficient; we will not need to perform the
remaining integrations.}
It is easy to check that
a calculation performed by closing the integration
contour in the upper real plane leads to a result identical to
\eqref{bubfin}. We will now re-derive this result using
the Feynman Tree Theorem.

\subsubsection{The bubble reloaded:
the Feynman Tree Theorem at work }

The starting point is the identity
\beq
\label{bubble-ret}
\cB_R (K^2) \ := \
\int\!{d^{D}L\over (2\pi)^{D}} \ \D_R (L) \, \D_R (L+K)
\ = \ 0
\ ,
\eeq
where $\D_R$ are the retarded propagators of
\eqref{ret}.
The vanishing of the left hand side of \eqref{bubble-ret}
is an immediate consequence of the fact that all the poles
of the integrand in the complex $L_0$ plane
lie below the real axis --
specifically they are
at $\o_L - i\vare$, $-\o_L - i \vare$, $-K_0 - \o_L - i \vare$,
$-K_0 + \o_L - i \vare$.
Closing the contour above, we see that $\cB_R =0$.

We now use Feynman's decomposition
\eqref{decret} for the two retarded
propagators $\D_R(L)$ and $\D_R (L+ K)$,
\beqa
\label{11}
\D_F (L) & = & \D_R(L) \ + \
{\pi \over \o_L} \, \d (L_0 + \o_L)
\ ,
\\
\label{22}
\D_F (L+K) & = & \D_R(L+K) \ + \
{\pi \over \o_{L+K}} \, \d (L_0 +K_0 + \o_{L+K})
\ ,
\eeqa
and get
an equation similar to
\eqref{ftt}, namely
\beq
\label{feydecbub}
\cB_F \ = \cB_{\rm 1-cut} \,  + \,  \cB_{\rm 2-cut}
\ .
\eeq
This is the Feynman Tree Theorem for the particular
case of the bubble diagram.
The one-particle cut contribution
arises from picking either one or the other
of the two delta functions in
\eqref{11} or \eqref{22}, and is
\beq
\cB_{\rm 1-cut}\ = \
\cB_{\rm 1-cut}^{(a)} \, + \, \cB_{\rm 1-cut}^{(b)}
\ ,
\eeq
with
\beqa
\nonumber
\cB_{\rm 1-cut}^{(a)}& = &
\int\!{d^DL \over (2\pi)^D}
\
{i \pi \, \d (L_0 + \o_L)\over \o_L \, \o_{L+K}}
\, \,
\bigg[
{1\over (L_0 + K_0) - \o_{L+K}  + i \vare}\,  - \,
{1\over (L_0 + K_0) + \o_{L+K} -i\vare }
\bigg]
\ ,
\\ %\nonumber
\cr
\cB_{\rm 1-cut}^{(b)}& = &
\int\!{d^DL \over (2\pi)^D}
\
{i \pi \, \d (L_0+K_0  + \o_{L+K})\over \o_L \, \o_{L+K}}
\, \,
\bigg[
{1\over L_0  - \o_{L}  + i \vare}\,  - \,
{1\over L_0  + \o_{L} -i\vare }
\bigg]
\ ,
\eeqa
from which it follows that
\beqa
\nonumber
\cB_{\rm 1-cut}
 =
{i\over 4}
\int\!{d^{D-1} \vec{L} \over (2\pi)^{D-1}}
\
{1 \over \o_L \, \o_{L+K}}
\, \,
&\bigg[ &
{1\over -K_0 - \o_L - \o_{L+K} + i \vare}\,  - \,
{1\over -K_0 + \o_L + \o_{L+K} - i \vare}
\\ [8pt]
&-& 2 \pi i \, \delta (\o_{L+K}  - \o_L + K_0 )
\bigg]
\ .
\label{1-cut-bubble}
\eeqa
Notice that \eqref{1-cut-bubble}
is equal to the result for $\cB_F$ obtained in \eqref{bubfin}
with standard methods except for an extra delta function
contribution.
In order to have agreement with
\eqref{feydecbub}, this additional term must be cancelled by
$\cB_{\rm 2-cut}$.
Indeed, a direct calculation of this term gives
\beqa
\nonumber
\cB_{\rm 2-cut} & = &
-\int\!{d^{D} L \over (2\pi)^{D}}
\
{\pi \over \o_L} \,
{\pi \over \o_{L+K}}
\ \d (L_0 + \o_L )
\,
\d (L_0 +K_0 + \o_{L+K} )
 \\ [8pt]\cr
 &=&
 - {\pi \over 2}
\int\!{d^{D-1} \vec{L} \over (2\pi)^{D-1}}
\ {\d (K_0 - \o_L + \o_{L+K} ) \over
\o_L \, \o_{L+K}}
\ ,
\label{supp}
 \eeqa
and it is immediately seen that \eqref{feydecbub}
holds.

In conclusion, we have seen that the application of
the Feynman Tree Theorem expressed by
\eqref{feydecbub} correctly  reproduces the
expected result \eqref{bubfin} for the bubble integral.

\subsubsection{A comment on Feynman two-particle cuts
and unitarity cuts}
Before concluding this section we would like to make
a comment on the Feynman two-particle cuts,
i.e.~the two-particle cuts appearing when using
the Feynman Tree Theorem,
such as \eqref{supp}. It is important to realise that
these are not the same as the conventional
unitarity cut%
\footnote{We mention however an interesting connection between
the Feynman Tree Theorem and Cutkosky rules, described in
section 7-3-3 of \cite{zuzu'}.}.
The reason is that the Feynman Tree Theorem forces the sign
of the time component of the two loop momenta in
Figure 16 to be the same, specifically
negative (positive)  when using the decomposition
of the Feynman propagator into
a retarded (advanced) propagator and a delta function.
The consequence is that this Feynman two-particle cut
vanishes when  $K^2 > 0$,
\beq
\label{2zero}
\cB_{\rm 2-cut}  \ = \ 0 \ ,
\qquad \quad {\rm if} \ \ K^2 > 0
\ ,
\eeq
therefore $\cB_{\rm 2-cut} $ cannot be a conventional
unitarity cut, which does not vanish above
the pair production threshold $K^2 > 0$.

To prove \eqref{2zero}, we notice that for  $K^2 > 0$
we can choose a frame where
$K = (K_0, \vec{0})$.
Then $\o_L = \o_{L+K}$, and the delta function
in the last expression in \eqref{supp} has support
for $K_0 = 0$; hence $K$ is vanishing, contradicting
the assumption $K^2 > 0$.
On the other hand, if  $K^2 < 0$ we can choose
a frame where $K = (0, \vec{K})$.
The delta function in \eqref{supp} now has support
for $\o_L = \o_{L+K}$, which admits the solution
$ \vec{L} = -\vec{K} / 2 $
(if $K^2 = 0$, dimensional regularisation
requires that $\cB_F (K^2 = 0 )$ must vanish).

%%%%%%%%%%%%%%%%%%%%%%%%%%%%%%%%%%%%%%%%%%%%%%%%%%%%%%%%%%%%%%%%

              \subsection{Cutting one-loop MHV amplitudes
                   with the Feynman Tree Theorem}

%%%%%%%%%%%%%%%%%%%%%%%%%%%%%%%%%%%%%%%%%%%%%%%%%%%%%%%%%%%%%%%%

Here we will apply the Feynman Tree Theorem to
the one-loop calculation of an MHV scattering amplitude.

The relevant diagram is shown in Figure 17.
From the point of view of the Feynman Tree Theorem,
this calculation is similar to that for
the bubble performed in the previous section. 
As in that case, Feynman's theorem is used to
decompose the two propagators
which enter the one-loop integral, and
the locality of the interaction makes it possible
to apply the theorem in both approaches.
The difference between the two calculations
lies in the parametrisation of the
loop momentum which is integrated over,
as we discuss now.

\subsubsection{The one-loop integration measure of \cite{bst}}

To begin with, we will re-derive the integration measure
which was used in \cite{bst} in order to calculate
one-loop MHV scattering amplitudes with MHV vertices.
The final result of this calculation, given in \eqref{bst},
will then be obtained in an alternative way by making use of
the Feynman Tree Theorem.

We choose a specific
parametrisation for a generic off-shell momentum
$L$ \cite{Bena:2004ry,Kosower:2004yz},
\beq
\label{off}
L \ = \ l \, + \, z \eta \ ,
\eeq
where $l^2=0$, and $\eta$ is a fixed and arbitrary
null vector, $\eta^2=0$; $z$ is a real number
(in real Minkowski space).
This choice of variables turns out to be particularly
convenient for calculating one-loop amplitudes with
MHV vertices \cite{bst}.
Using \eqref{off}, one can solve for $z$
as a function of $L$,
\beq
z \ = \
{L^2 \over  2 (L \eta)}
\ .
\eeq
Using spinor notation, we  write $l$ and $\eta$
as $l_{\a \da} = l_{\a} \llt_{\da}$,
$\eta_{\a \da} = \eta_{\a} \tilde{\eta}_{\da}$.
It then follows that
\beqa
\label{1}
l_\a & = & {L_{\a \da} \tilde{\eta}^{\da}
\over [ \llt \, \tilde{\eta}]
}
\ ,
\\
\label{2}
\llt_{\da} & = & {\eta^\a L_{\a \da} \over \lan l \, \eta
\ran} \ .
 \eeqa
\eqref{1} and \eqref{2} are equivalent
to the CSW prescription proposed in \cite{csw}
to determine the spinor variables
$l$ and $\llt$ associated with the non-null,
off-shell four-vector $L$ defined in \eqref{off}.%
\footnote{The denominators on the right
hand sides of \eqref{1} and \eqref{2} are
irrelevant for applications, since the expressions
we will be dealing with are homogeneous in the spinor
variables $l_{\alpha}$.}

In \cite{bst} the integration
measure $d^4L$ over a generic loop momentum $L$
was re-expressed in terms
of the new variables $l$ and $z$ introduced previously,
with the result%
\footnote{Here we define $d^4L := \prod_{i=0}^{3} d L_i$.}
\beq
\label{leporello}
{d^4 L \over
L^2} \ = \
{dz \over z } \, {d\cN (l)}
\ ,
\eeq
where we have introduced the Nair measure  \cite{Nair}
\beq\label{nairmeas}
 d\cN (l) \ := \ {1\over 4i} \Big(
 \lan l
\, \, dl \ran \, d^2 \tilde{l} \ - \ [ \tilde{l} \, \, d
\tilde{l}] \, d^2 l
\Bigr)   = {d^3 l \over 2\, l_0}\ ,
\eeq
and $l_0 = (1 / 2) ( l_1 \tilde{l}_{\dot{1}} +
l_2 \tilde{l}_{\dot{2}} )$ is the time component of the
on-shell four-vector $l:= (l_0, \vec{l})$.
It is important to realise that \eqref{nairmeas}
does not contain any step function requiring the sign
of the time-component $l_0$ of $l$ to be positive or negative;
both possibilities
$l_0  = \pm |\vec{l}|$ are allowed.%
\footnote{In Minkowski space we identify
$\tilde{l} = \pm l^{\ast}$ depending on whether
$l_0$ is positive or negative.
 }
From \eqref{leporello} it follows that
\beq
\label{commendatore}
d^4L \ = \
dz \, {d^3l \over 2\, l_0} \, 2 (l\eta)
\ .
\eeq
Since both $l$ and $\eta$ are null it follows that
${\rm sgn} (l \eta) ={\rm sgn} (l_0 \eta_0)$
($\eta_0$ is  the time component of  $\eta$),
so that the Jacobian in \eqref{commendatore}
is positive if $\eta_0>0$.
We will assume this throughout this paper.
Finally, we notice that
\beq
\label{mmm}
{d^4 L \over
L^2 + i \varepsilon} \ = \
{dz \over z   \, + \, i\,  {\rm sgn } (l_0 \eta_0) \varepsilon }
\, {d\cN (l)}
\ ,
\eeq
with $d\cN (l)$ given by \eqref{nairmeas}.

In computing one-loop MHV scattering amplitudes
from MHV diagrams (shown in Figure 17),
the four-dimensional integration measure
is \cite{bst}
\beq
\label{dem}
d\cM \ := \ {d^4 L_1 \over L_1^2 + i \varepsilon}
\,
{d^4 L_2 \over L_2^2 + i \varepsilon}
\
\delta^{(4)} (L_2 - L_1 + P_L)
\ ,
\eeq
where $L_1$ and $L_2$ are loop momenta, and $P_L$ is the
external momentum flowing outside the loop%
\footnote{In our conventions, all external momenta are
outgoing.}
so that $L_2 - L_1 + P_L=0$.
Now we express  $L_1$ and $L_2$ as in
\eqref{off},
\beq
\label{ells}
 L_{i; \a, \da} \ =
\ l_{i \a} \llt_{i \da}  \, + \,
z_i \, \eta_{\a}\tilde{\eta}_{\da} \ ,
\qquad i=1,2 \ .
 \eeq
Using \eqref{ells},
we  rewrite the argument of the delta function as
\beq
L_2 - L_1 + P_L
= l_2 - l_1 + P_{L; z} \ ,
\eeq
where
\beq
\label{plz}
P_{L;z} := P_L - z \eta
\ ,
\eeq
and
\beq
\label{zeta}
z \ := \ z_1\, - \, z_2
\ .
\eeq
Notice that we use the same $\eta$ for both the momenta
$L_1$ and $L_2$.
Using \eqref{ells}, we can re-cast \eqref{dem} as
\cite{bst}
\beq
\label{gcar}
d\cM \ = \
{dz_1 \over z_1 + i \vare_1}\, { dz_2 \over z_2 + i \vare_2}
\
\bigg[ {d^3 l_1 \over 2 l_{10}}\,
{d^3 l_2 \over 2 l_{20}}
\,
\delta^{(4)}  (l_2 \, - \, l_1 \,  + \, P_{L; z} )
\bigg]
\ ,
\eeq
where
$\vare_{i} := {\rm sgn}(\eta_0 l_{i0}) \vare =
{\rm sgn}(l_{i0})\vare$, $i = 1, 2$ (the last equality
holds since we are assuming $\eta_0 >0$).
Notice also that no step functions for $l_{10}$ or $l_{20}$
have appeared so far.

\begin{figure}[ht]
\label{Figurelast}
\begin{center}
\scalebox{0.75}{\includegraphics{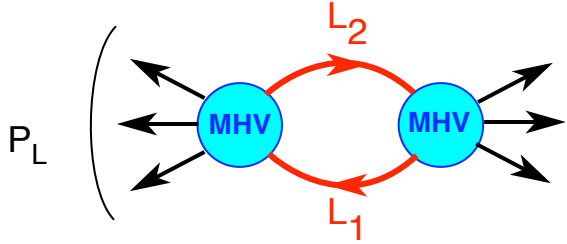}}
\end{center}
\caption{\it MHV diagrams contributing to a
one-loop MHV scattering amplitude.
In the text we show that these diagrams can
equivalently be computed either by a direct calculation,
as in \cite{bst}, or by resorting to the Feynman
Tree Theorem.
}
\end{figure}

We can now convert the integration over $z_1$ and $z_2$
in \eqref{gcar} to an integration over $z$ defined in \eqref{zeta},
and $z'$ defined by
\beq
z' \ :=  z_1 \, + \, z_2
\ .
\eeq
Indeed
\cite{bst}, neither the measure nor the
integrand depend on $z'$,
which can therefore be integrated out.
To do this, we observe that
\beq
{dz_1 \over z_1 + i \vare_1}\, { dz_2 \over z_2 + i \vare_2}
\ \rightarrow \
2\, {dz \, dz'  \over (z' + z + i \vare_1)(z' - z + i \vare_2)
}
\ .
\eeq
The previous expression viewed as a function of
$z'$  has two simple poles,
located at $-z-i\vare_1$ and $ z - i \vare_2$.
To carry out the $z'$ integration, we must
discuss in detail the $i \vare$ prescriptions.

We begin by  first assuming that
$P_{L;z}^2 > 0$.
We can then go to the rest frame of
$P_{L;z}$, where $P_{L;z} = a (1, \vec{0})$ for some real
number $a$.
Momentum conservation requires
\beq
\label{a-s}
l_1 \ = \ {a \over 2} (1 , \hat{n}) \ ,
\qquad \quad
l_2 \ = \ {a \over 2} (-1 , \hat{n}) \ ,
\eeq
where $\hat{n}$ is a unit vector.
Note that the time components of $l_1$ and
$l_2$ have opposite signs.
These signs are determined by the sign of $a$ or,
stated in an invariant way,
by the sign of the time component $(P_{L;z})_0$ of
$P_{L;\, z}$ (recall that the sign of the time component
of a time-like vector is a proper Lorentz invariant).
We can furthermore make the replacement
\beq
\label{defini}
 {d^3 l_1 \over 2 l_{10}}\,
{d^3 l_2 \over 2 l_{20}}
\,
\delta^{(4)}  (l_2 \, - \, l_1 \,  + \, P_{L; z} )
\ \rightarrow  \
- \, d{\rm LIPS}  (l_2^{-} , - l_1^{+}; P_{L;z})
\ ,
\eeq
where
\beq
\label{LIPS}
d{\rm LIPS} (l_2^{-}  , - l_1^{+} ; P_{L;z}) \ := \
d^4 l_1 \, \delta^{(+)} (l_1^2) \
d^4 l_2 \, \delta^{(-)} (l_2^2 )\
\delta^{(4)} (l_2 - l_1 + P_{L;z})
\
\eeq
is the two-particle Lorentz invariant phase
space  (LIPS) measure, and we
recall that $ \delta^{\pm} (l^2) : =
\theta ( \pm l_0 ) \delta (l^2)$.

It is crucial that at $P_{L;z}^2>0$,
$\vare_1$ and $\vare_2$ have opposite signs.
By closing the integration contour with a large semicircle
in the upper (or lower) $z'$ complex plane,
one always picks the contribution of one pole.
We first assume the sign of the time component
of $P_{L;z}$ to be positive,
$a>0$.
Remember that $\vare_i := {\rm sgn} (l_{i0} \eta_0)$;
having assumed $\eta_0 > 0$, we see that
\beq
(P_{L;z})_0 > 0 \ \ \Rightarrow \ \ \vare_1 > 0 \ ,
\qquad \vare_2 < 0
\ .
\eeq
Performing the $z'$ integration by closing the contour
either above or below one gets
\beq
2\, {dz \, dz'  \over (z' + z + i \vare_1)(z' - z + i \vare_2)}
\ \rightarrow \
 2 \pi i \, {dz \over z + i \vare}
\ .
\eeq
Therefore from \eqref{gcar} we get
\beq
\label{replace2}
\left. d\cM\right|_{P_{L;z}^2 > 0; \, (P_{L;z})_0 > 0 } \ = \
-\,  2 \pi i \ {dz \over z + i \vare}
\
d{\rm LIPS} (l_2^{-} , - l_1^{+}; P_{L;z})
\ .
\eeq
What happens if
we choose $a<0$, i.e.~$(P_{L;z})_0 <0$?
There are two differences compared with the case above.
The first one is
\beq
(P_{L;z})_0 < 0 \ \ \Rightarrow \ \ \vare_1 < 0 \ ,
\qquad \vare_2 > 0
\ ,
\eeq
and the result of the $z'$ integration is now
\beq
2\, {dz \, dz'  \over (z' + z + i \vare_1)(z' - z + i \vare_2)}
\ \rightarrow \
 2 \pi i \, {dz \over - z \, + \,  i \vare}
\ .
\eeq
Secondly, we have now
\beq
 {d^3 l_1 \over 2 l_{10}}\,
{d^3 l_2 \over 2 l_{20}}
\,
\delta^{(4)}  (l_2 \, - \, l_1 \,  + \, P_{L; z} )
\ \rightarrow  \
- \,
d{\rm LIPS} (l_2^{+} , - l_1^{-}; P_{L;z})
\ .
\eeq
Hence one would obtain
\beq
\label{replace2bis}
\left. d\cM\right|_{P_{L;z}^2 > 0; \, (P_{L;z})_0 < 0 } \ = \
-\, 2 \pi i \ {dz \over - z \, + \, i \vare}
\
d{\rm LIPS} (l_2^{+} , - l_1^{-}; P_{L;z})
\ .
\eeq
We turn to the case where $P_{L;z}^2 <0$.
In this case one can transform  to a system
where $P_{L;z} = (0, \vec{\a})$ for some three vector
$\vec{\a}$.
Then momentum conservation requires
\beq
l_1 = b  (1, \hat{m}_1)
\ ,
\qquad \quad
l_2 = b  ( 1 , \hat{m_2})
\  ,
\eeq
with
$b (\hat{m}_2-\hat{m}_1 ) + \vec{\a} = 0$.
Now, the sign of the time components of
$l_1$ and $l_2$ are the same.
Following steps similar to above, and noticing
that now it is possible to close the integration contour
in $z'$ without encircling  any poles,
we find  that
\beq
\left. d\cM\right|_{P_{L;z}^2  < 0} \ = \
0
\ .
\eeq
Therefore, we have re-cast
the one-loop integration measure \eqref{gcar}
as
\beq
\label{final}
d\cM \ = \
-\,  2 \pi i \
\theta (P_{L;z}^2)
\ {dz \over z \, {\rm sgn} (P_{L; z})_0 \, + \, i \vare}
\
d{\rm LIPS} (l_2^{\mp} , - l_1^{\pm}; P_{L;z})
\ ,
\eeq
where the upper (lower) sign in
the superscripts is for $(P_{L; z})_0 > 0$
($(P_{L; z})_0 < 0 $).
An alternative way to write the measure is
\beq
\label{superfinal}
d\cM \ = \
-\,  2 \pi i \
\theta (P_{L;z}^2)\
{dz \over z \, {\rm sgn} (l_{10}) \, + \, i \vare}
\
d{\rm LIPS} (l_2^{\mp} , - l_1^{\pm}; P_{L;z})
\ ,
\eeq
where the upper (lower) sign is for $l_{10} > 0 $
($l_{10} < 0 $).

Finally, we observe that
one can trade the $z$ integration for an
integration over $P_{L;z}^2$.
After a little algebra, one finds that
for $P_{L;z}^2 >0$
\beq
{dz\over z \, {\rm sgn } \, (P_{L;z})_0 \, + \, i \vare } \ = \
{dP_{L;z}^2 \over - P_{L;z}^2 \, +  \, P_L^2 \, + \, i \vare}
\ .
\eeq
In conclusion, the integration measure \eqref{final}
gives
\beq
\label{bst}
d\cM \ = \
 2 \pi i \
\theta (P_{L;z}^2)\
{dP_{L;z}^2 \over P_{L;z}^2 \, - \, P_L^2 \, - \, i \vare}
\
d{\rm LIPS} (l_2^{\mp} , - l_1^{\pm}; P_{L;z})
\ ,
\eeq
which is our final result.

A few comments are  in order here.

{\bf 1.}
As \eqref{bst} shows,
the integration is performed for
$P_{L;z}^2 > 0$. By setting all the various external
kinematical invariants $P_{L}^2$ to be negative,
no poles are encountered along the integration contour,
and the $i \vare$ prescription can be dropped.
This is what had been done in \cite{bst}.
However \eqref{bst} provides
us with the correct analytic continuation to
the physical region, which is obtained by simply performing
the substitution
\beq
\label{repp}
P^2_L \ \longrightarrow \ P^2_L \, + \, i \vare
\ .
\eeq
In \cite{bst} a  new form of the
two-mass easy box function was derived,
simpler than the usual form.
One of the advantages of this
expression  was precisely that
its analytic continuation in a physical region
is achieved by simply performing the
naive replacement dictated by \eqref{repp}.

{\bf 2. }
A comment may be made on the phase space measure appearing in
\eqref{bst}. This instructs us to include both possibilities
$l_{10} = |\vec{l_1}|$, $l_{20} = -|\vec{l_2}|$,
and $l_{10} = -|\vec{l_1}|$, $l_{20} = |\vec{l_2}|$.
We can swap the sign of $l_0$ by simply replacing
$\tilde{l} \to - \tilde{l}$, where $l_{\a \dot{\a}} :=
l_\a \tilde{l}_{\dot{\a}}$.
MHV vertices do not contain dotted spinor variables,
hence it is not necessary to distinguish between the
cases $l_{10} > 0$ and $l_{10} < 0$.
Taking this into account, we can simply
choose e.g.~$l_{10} > 0$ and multiply the final result
by a factor of two.

{\bf 3.}
The integration measure $d\cM$ as it is expressed
on the right hand side of \eqref{bst}
can immediately be dimensionally regularised,
by simply replacing the four-dimensional LIPS measure
of
\eqref{LIPS} by
its continuation to $D=4-2\epsilon$ dimensions,
\beq
\label{LIPSD}
d^D{\rm LIPS} (l_2^- , - l_1^+; P_{L;z}) \ := \
d^D l_1 \, \delta^{(+)} (l_1^2) \
d^D l_2 \, \delta^{(-)} (l_2^2 )\
\delta^{(D)} (l_2 - l_1 + P_{L;z})
\ .
\eeq

{\bf 4.}
Eq.~\eqref{bst}
was  one of the key results of  \cite{bst}.
It gives a decomposition of the original
integration measure into a phase space measure and a
dispersive measure.
From Cutkosky's theorem \cite{Cutkosky:1960sp}
we know that the LIPS measure computes
the discontinuity of a Feynman diagram across
its branch cuts. Which discontinuity is evaluated
is determined by the argument of the delta function
appearing in the LIPS measure;
in \eqref{bst} this is $P_{L;z}$
(defined in \eqref{plz}).
Finally, discontinuities are integrated using the
dispersive measure in \eqref{bst},
thereby reconstructing the full amplitude.

\subsubsection{Re-derivation of the measure of \cite{bst}
 with the Feynman Tree Theorem}

Here we show that the integration measure \eqref{bst}
can equivalently be derived by applying the Feynman Tree Theorem
directly to the measure \eqref{gcar}.
This provides a check of the consistency of our prescriptions.

We start off by considering the Feynman one-particle cuts.
When we cut the leg with momentum $L_2$ we replace it by its
on-shell version with momentum $l_2$, and the integration
measure $d^4 L_2 / (L_2^2 + i \vare)$ is replaced by what
is prescribed by the second term in \eqref{decret};
dividing by a factor of $i$ (since in \eqref{gcar}
we had omitted factors of $i$ in the propagators),
we have from \eqref{ftt} that the once-cut measure
is obtained by making the replacement
\beq
\label{donnaelvira}
{ d^4 L_2 \over  L_2^2 + i \vare }
\ \rightarrow \ - 2 \pi i \ d^4L_2 \, \delta^{(-)} (L_2^2)
\ = \
2 \pi i \,
\left.
{d^3 l_2 \over 2 l_{20}}
\right|_{l_{20} = - |\vec{l_2}|}
\ .
\eeq
The once-cut contribution to the integration measure
arising from cutting $L_2$ is therefore
\beq
\label{gcar2}
\left.
d\cM
\right|_{L_2-{\rm cut}}  \ = \
2\pi i \ {dz_1 \over z_1 + i \vare_1}\,
\
\bigg[ {d^3 l_1 \over 2 l_{10}}\,
{d^3 l_2 \over 2 l_{20}}
\,
\delta^{(4)}  (l_2 \, - \, l_1 \,  + \, P_{L; \, z_1} )
\bigg]_{l_{20} < 0}
\ .
\eeq
On the other hand, if we cut $L_1$ we get
\beq
\label{gcar1}
\left.
d\cM
\right|_{L_1-{\rm cut}}  \ = \
2\pi i \ {dz_2 \over z_2 + i \vare_2}\,
\
\bigg[ {d^3 l_1 \over 2 l_{10}}\,
{d^3 l_2 \over 2 l_{20}}
\,
\delta^{(4)}  (l_2 \, - \, l_1 \,  + \, P_{L; \, - z_2} )
\bigg]_{l_{10} < 0}
\ .
\eeq
Changing variable in \eqref{gcar1}
from $z_2$ to $-z_2$, we get
\beq
\label{gcar11}
\left.
d\cM
\right|_{L_1-{\rm cut}}  \ = \
 2\pi i \
 {dz_2 \over -\, z_2 + i \vare_2}\,
\
\bigg[ {d^3 l_1 \over 2 l_{10}}\,
{d^3 l_2 \over 2 l_{20}}
\,
\delta^{(4)}  (l_2 \, - \, l_1 \,  + \, P_{L; \,  z_2} )
\bigg]_{l_{10} < 0}
\ .
\eeq
It is also useful to rename $z_{1,2} \to z$ in the previous
formulae.

Now, for $P_{L; z}^2 >0$ we know that the signs
of the time component of $l_1$ and $l_2$ are opposite.
Remembering that $\vare_i = {\rm sgn } (l_{i0})\vare $
(having assumed $\eta_0 > 0$), we see that
$\vare_1 > 0$ in \eqref{gcar2}, and
$\vare_2 >0$ in \eqref{gcar11}.
In this case, \eqref{gcar2} becomes
\beq
\label{gcar2-bis}
\left.
d\cM
\right|_{L_2-{\rm cut}; P_{L;z}^2>0}  \ = \
- 2\pi i \ {dz \over z + i \vare}\,
\
d{\rm LIPS} (l_2^-, -l_1^+; P_{L;z})
\ ,
\eeq
whereas \eqref{gcar11} gives
\beq
\label{gcar11-bis}
\left.
d\cM
\right|_{L_1-{\rm cut}; P_{L;z}^2>0}  \ = \
 -2\pi i \
 {dz \over -\, z + i \vare}\,
\
d{\rm LIPS} (l_2^+, -l_1^-; P_{L;z})
\ .
\eeq
For $P_{L;z}^2 >0$, \eqref{gcar2-bis} and \eqref{gcar11-bis}
are already in agreement with the  measure
\eqref{final} (or \eqref{superfinal});
hence we expect that, in this regime, the two-particle cuts
should not contribute. This would indeed completely parallel the
example of the bubble addressed in an earlier section.

But before moving to the two-particle cuts, let us
complete the study of the Feynman one-cuts, by considering the
regime $P_{L:z}^2 < 0$.
Here momentum conservation requires that,
both in \eqref{gcar2} and \eqref{gcar11},
one must have $l_{10} <0 $ and $l_{20} < 0$.
Hence $\vare_1$ and $\vare_2$ have the same (negative) sign.
It follows that at $P_{L;z}^2 < 0$ the net effect of \eqref{gcar2}
and \eqref{gcar11} is given by a common
phase space measure, multiplied by
\beq
\label{ultimo}
{1\over z- i \vare} + {1\over -z  - i \vare} \ = \
 2 i \pi \delta (z)
\ .
\eeq
Integrating out $z$ using the delta function, one obtains
\beq
\label{zerlinetta}
\left.
d\cM_{P_{L;z}^2 < 0}^{1-{\rm cut}} \ = \
(2\pi i )^2
\
{d^3 l_1 \over 2 l_{10}}\,
{d^3 l_2 \over 2 l_{20}}
\,
\delta^{(4)}  (l_2 \, - \, l_1 \,  + \, P_{L} )
\right|_{l_{10} <0,\, l_{20} < 0}
\ .
\eeq
We know that the measure \eqref{bst}
vanishes for $P_{L;z}^2 < 0$; hence, by the Feynman Tree Theorem
we expect that a two-particle cut contribution will cancel
\eqref{zerlinetta}.

Indeed the Feynman two-particle cut is precisely nonvanishing
when $P_{L;z}^2 <0$.
By replacing the two propagators
in the loop with  two delta functions
similarly to \eqref{donnaelvira},
we immediately get
\beq
d\cM_{P_{L;z}^2 < 0}^{2-{\rm cut}} \ = \
-\, (2\pi i )^2
\
{d^3 l_1 \over 2 l_{10}}\,
{d^3 l_2 \over 2 l_{20}}
\,
\delta^{(4)}  (l_2 \, - \, l_1 \,  + \, P_{L} )
\ ,
\eeq
which precisely cancels
\eqref{zerlinetta},
\beq
d\cM_{P_{L;z}^2 < 0}^{1-{\rm cut}} \ + \
d\cM_{P_{L;z}^2 < 0}^{2-{\rm cut}} \ = \
0
\ .
\eeq
On the other hand, when $P_{L;z}^2>0$
the Feynman two-particle cut vanishes.
We had already observed that, for
$P_{L;z}^2 > 0$, the measure from
one-particle cuts reproduces the
measure of \cite{bst}, therefore we
conclude that the Feynman Tree Theorem
precisely reconstructs
the measure \eqref{bst}
of \cite{bst}.

%%%%%%%%%%%%%%%%%%%%%%%%%%%%%%%%%%%%%%%%%%%%%%%%%%%%%%%%%%%%%%%%

\section{Conclusions}

%%%%%%%%%%%%%%%%%%%%%%%%%%%%%%%%%%%%%%%%%%%%%%%%%%%%%%%%%%%%%%%%

In this paper we have argued that any one-loop 
scattering amplitude in supersymmetric Yang-Mills theories 
can be computed with MHV diagrams. 
This was based upon the following.

Firstly, we have shown that all one-loop MHV diagrams produce 
results which are covariant, i.e.~independent of 
the reference spinor introduced to define the 
off-shell continuation of MHV amplitudes to vertices. 
The key ingredient in this proof is the Feynman Tree
Theorem -- a simple result, but a powerful one in that 
it allows one to infer properties of loops from those of 
trees. We comment that its applicability 
extends to massive particles, as well as to higher loops. 

Next, we have shown that MHV diagrams give amplitudes with the 
correct cuts, and correct generalised cuts, 
in all kinematical channels.
We then studied collinear and soft limits 
of scattering amplitudes. 
In particular, using MHV diagrams we derived 
the universal behaviour of amplitudes
in limits where two momenta become collinear. 
The corresponding one-loop gluon splitting functions 
were derived to all orders in 
the dimensional regularisation parameter, $\e$, 
and found to be in perfect agreement with 
the results of \cite{ku,vittorio}. 
This is rather remarkable, and we believe that 
it supports the expectation that the
MHV diagram method might be valid beyond one loop.

In the last section of the paper 
we have presented applications 
of the Feynman Tree Theorem to one-loop Feynman and MHV diagrams. 
In particular, we have given an efficient re-derivation 
of the one-loop integration measure of \cite{bst}
using the Feynman Tree Theorem. 

In order to complete the full proof of 
the equivalence of the Feynman diagram 
and MHV diagram methods, 
there remains the issue of the factorisation 
on multi-particle poles of one-loop scattering amplitudes.
This should follow using similar arguments 
to those presented here.
We also expect that the techniques 
discussed in this paper will be useful in furthering 
our understanding of one- and multi-loop 
scattering amplitudes in gauge theories, perhaps 
including theories with massive particles.

%%%%%%%%%%%%%%%%%%%%%%%%%%%%%%%%%%%%%%%%%%%%%%%%%%%%%%%%%%%%%
%%%%%%%%%%%%%%%%%%%%%%%%%%%%%%%%%%%%%%%%%%%%%%%%%%%%%%%%%%%%%

                \section*{Acknowledgements}

%%%%%%%%%%%%%%%%%%%%%%%%%%%%%%%%%%%%%%%%%%%%%%%%%%%%%%%%%%%%%
%%%%%%%%%%%%%%%%%%%%%%%%%%%%%%%%%%%%%%%%%%%%%%%%%%%%%%%%%%%%%

It is a pleasure to thank Michael Green for bringing
the Feynman Tree Theorem to our attention,
and Adi Armoni, Valya Khoze, Marco Matone
and Sanjaye Ramgoolam for discussions.
We would like to thank the University of Durham 
for hospitality during the London Mathematical Society 
symposium on {\it Geometry, Conformal Field Theory and String Theory},
where this work was started.
AB and GT thank  the Theory Division at CERN
for hospitality and support at various stages 
of this work. This work was partially supported
by a Particle Physics and Astronomy Research Council 
award {\it M Theory, string theory and duality}.
The work of GT is supported by an
EPSRC Advanced Fellowship.

\newpage

%%%%%%%%%%%%%%%%%%%%%%%%%%%%%%%%%%%%%%%%%%%%%%%%%%%%%%%%%%%%
\startappendix
%%%%%%%%%%%%%%%%%%%%%%%%%%%%%%%%%%%%%%%%%%%%%%%%%%%%%%

\Appendix{All-order in $\e$ expressions for
the two-mass easy box functions}
In this appendix we present
five all-order in $\e$ expressions for the
two-mass easy box function
$F^{\rm 2me} (s, t, P^2, Q^2)$, where $p$ and $q$
are the massless legs, with 
$s:= (P+p)^2$, $t:= (P+q)^2$.
It is convenient to define the variable
$u:= (p+q)^2$, with $s+t+u = P^2 + Q^2$, and
\beq
a \ := \ {u \over P^2 Q^2 - st} \ = \
{2 (pq) \over P^2 Q^2 - st}
\ .
\eeq
The first form is
\beqa
\nonumber
%\hspace{-3cm}
&& \hspace{-0.7cm}
F^{\rm 2me} (s, t, P^2, Q^2) =
-{ c_{\G} \over \e^2}
\left[
 \Big( {-s \over \mu^2} \Big)^{-\e} \,
\mbox{}_{2}F_1 \left( 1, -\e, 1- \e, as \right)
\, + \,
\Big( {-t \over \mu^2} \Big)^{-\e}
\mbox{}_{2}F_1 \left( 1, -\e, 1- \e, at \right)
\right.
\\  [6pt]\cr
&&  \qquad - \,
\left.\Big( {-P^2 \over \mu^2} \Big)^{-\e}
\,
\mbox{}_{2}F_1 \left( 1, -\e, 1- \e, aP^2 \right)
\, - \,
 \Big( {-Q^2 \over \mu^2} \Big)^{-\e} \,
\mbox{}_{2}F_1 \left( 1, -\e, 1- \e, a Q^2 \right)
\right]
\ ,
\label{puzzola}
\eeqa
where, as usual,
\beq
c_\G \ := \ {\G (1 + \e) \G^2 ( 1 -  \e) \over (4\pi)^{2- \e}
\G(1 - 2 \e)}
\ .
\eeq
We have explicitly used this expression 
in section 4 and explained in footnote 16 how this form is
obtained from a generalisation of 
a computation in \cite{bst}. Notice that
\beq
\mbox{}_{2}F_1 \left( 1, -\e, 1- \e, z \right)
\ = \ -\e \, z^\e \, B_{z} ( -\e ,  0)
\ ,
\eeq
where $B_z (a,b)$ is the incomplete beta function.
Hence we can rewrite $F^{\rm 2me} (s, t, P^2, Q^2)$
in the following compact form:
\beq
%\hspace{-3cm}
%&& \hspace{-0.7cm}
F^{\rm 2me} (s, t, P^2, Q^2) =
{ c_{\G} \over \e} ( - a \m^2 )^\e
\Big[
 B_{as} ( -\e, 0) \, + B_{at} ( -\e, 0)
\, -  \, B_{aP^2} ( -\e, 0) \, -  B_{aQ^2} ( -\e, 0)
\Big]
.
\label{puzzoletta}
\eeq

Let us now find the four-dimensional limit of \eqref{puzzola}.
Using the expansion
\beq
\mbox{}_{2}F_1 \left( 1, -\e, 1- \e, z \right)
\ = \ 1 \, - \,
\sum_{m=1}^{\infty} \e^m \,
{\rm Li}_{m} (z)
\ ,
\eeq
for each of the  four hypergeometric functions
in \eqref{puzzola},
we see that the logarithmic term from the
$\cO (\e)$ term in the expansion of
the hypergeometric functions cancels due to the identity
\cite{bst}
\beq
\label{bah}
(1-as)(1-at) \ =  \
(1-aP^2)(1-aQ^2)
\ .
\eeq
At  $\cO (\e^2)$ we use
four times Euler's identity
\beq
\label{dilogids}
\nonumber
-{\rm Li}_2 (x) \, - \,
\log(x)\log(1-x) \ = \ {\rm Li}_2(1-x) \, - \, {\pi^2\over 6}
\ ,
\eeq
as well as \eqref{bah},
to find that,
up to terms which vanish when $\e \to 0$,
$F^{\rm 2me} (s, t, P^2, Q^2) $ becomes
\beqa
\label{niceonecyril}
\nonumber
 F^{\rm 2me} (s,t,P^2, Q^2) &=&
c_\Gamma \bigg\{ -\frac{1}{\e^2}\bigg[ \left(-s\over \mu^2\right)^{-\e} \, + \,
\left({-t\over \m^2}\right)^{-\e} \, - \,
  \left({-P^2\over \m^2}\right)^{-\e}\, - \,
\left({-Q^2\over \m^2}\right)^{-\e}\bigg]  
\\ [6pt]\nonumber  \cr
 &+&
\Li2(1-aP^2)\, + \, \Li2(1-aQ^2)  \, -\,  \Li2(1-as)
\,  -\,  \Li2(1-at) 
\bigg\}\ .
\\
\eeqa
This coincides with the form of the two-mass easy
box functions found in \cite{bst} 
using dispersion integrals.%
\footnote{We have recently become aware of 
the interesting paper \cite{dnizic}, 
which already contains the form of the
two-mass easy box function we obtained using 
dispersion integrals in \cite{bst}.
The paper \cite{dnizic} also discusses 
in detail the analytic continuation
to the physical region of the form \eqref{niceonecyril}
of the box function, which  is simpler than that 
required for the standard form of the 
same function (which appears e.g.~in \cite{ultimo}).}

Now we present four more forms of the all-order in $\e$
two-mass easy box functions.
These can be obtained from the previous form
by using various identities relating hypergeometric functions,
in particular
\beq
\label{v1}
\mbox{}_{2}F_1 \left( c-a, b, c, {z\over z-1} \right)
\ = \ (1-z)^b \mbox{}_{2}F_1 \left( a, b, c, z \right)
\ ,
\eeq
and
\beq
\label{v2}
\mbox{}_{2}F_1 \left( 1, \e, 1+\e, z^{-1}  \right)
\ = \ 1 \, - \, \mbox{}_{2}F_1 \left( 1, -\e, 1-\e, z \right)
\, + \, (-z)^{\e} \G( 1 + \e) \G (1 - \e)
\ .
\eeq
These  forms are:
\beqa
\nonumber
&& \hspace{-0.7cm}
F^{\rm 2me} (s, t, P^2, Q^2) \ =  \
-{ c_{\G} \over \e^2} \, (a \m^2)^{\e}
\\ [6pt]\nonumber
&&\hspace{-0.7cm} 
\times
\left[
\Big( {-as  \over 1-as} \Big)^{-\e} \,
\mbox{}_{2}F_1 \left( -\e, -\e, 1- \e, {-as \over 1 - a s }\right)
\, + \,
\Big( {-a t \over 1-at } \Big)^{-\e} \,
\mbox{}_{2}F_1 \left( -\e, -\e, 1- \e, {-at  \over 1-at  }\right)
\right.
\\ [6pt] \cr
&&\hspace{-1.2cm}  - \,
\left.
\Big( {-a P^2 \over 1-aP^2} \Big)^{-\e} \,
\mbox{}_{2}F_1 \left( -\e,- \e, 1- \e, {-aP^2  \over 1 - a P^2}\right)
\, - \,
 \Big( {-a Q^2\over 1-aQ^2} \Big)^{-\e} \,
\mbox{}_{2}F_1 \left( -\e, -\e, 1- \e, {-aQ^2 \over 1 - a Q^2 }\right)
\right], 
\nonumber \\
\label{444}
\eeqa
\beqa
\label{222}
\nonumber
&& \hspace{-1.2cm}
F^{\rm 2me} (s,t,P^2, Q^2) \ = \
- {c_\Gamma \over \epsilon^2}\bigg\{ \bigg[ \left(-s\over \mu^2\right)^{-\e} \, + \,
\left({-t\over \m^2}\right)^{-\e} \, - \,
  \left({-P^2\over \m^2}\right)^{-\e}\, - \,
\left({-Q^2\over \m^2}\right)^{-\e}\bigg]  \\ [6pt]\nonumber  \cr
&&\hspace{-.4cm}
- \Big( {-s \over \mu^2} \Big)^{-\e} \,
\mbox{}_{2}F_1 \left( 1, \e, 1+ \e, (as)^{-1} \right)
\, - \,
\Big( {-t \over \mu^2} \Big)^{-\e}
\mbox{}_{2}F_1 \left( 1, \e, 1+ \e, (at)^{-1} \right)
\\  [6pt]\cr
&& \hspace{-.4cm}
 +
\Big( {-P^2 \over \mu^2} \Big)^{-\e}
\,
\mbox{}_{2}F_1 \left( 1, \e, 1+ \e, (aP^2)^{-1} \right)
\, + \,
 \Big( {-Q^2 \over \mu^2} \Big)^{-\e} \,
\mbox{}_{2}F_1 \left( 1, \e, 1+ \e, (a Q^2)^{-1} \right)
\bigg\},
\nonumber \\
\eeqa
and
\beqa
\nonumber
%\nonumber
%\hspace{-3cm}
&& \hspace{-0.7cm}
F^{\rm 2me} (s, t, P^2, Q^2) =
-{c_\Gamma \over \e^2}
\left[
 \Big( {-s \over \mu^2} \Big)^{-\e} \,
\, + \,
\Big( {-t \over \mu^2} \Big)^{-\e}
\, - \, 
\Big( {-P^2 \over \mu^2} \Big)^{-\e}
\, - \, 
 \Big( {-Q^2 \over \mu^2} \Big)^{-\e} \,
\right. 
\\ 
&& \hspace{-0.3cm} \left. + \, 
\Big( {a \m^{2} \over 1-aP^2} \Big)^{\e} \,
\mbox{}_{2}F_1 \left( \e, \e, 1+ \e, {1 \over 1 - a P^2 }\right)
\, + \,
\Big( {a \m^{2} \over 1-aQ^2} \Big)^{\e} \,
\mbox{}_{2}F_1 \left( \e, \e, 1+ \e, {1 \over 1 - a Q^2 }\right)
\right.
\nonumber 
\\ [6pt] \cr
&& \hspace{-0.3cm}  - \,
\left.
\Big( {a \m^{2} \over 1-as} \Big)^{\e} \,
\mbox{}_{2}F_1 \left( \e, \e, 1+ \e, {1 \over 1 - a s }\right)
\, - \,
 \Big( {a \m^{2} \over 1-at} \Big)^{\e} \,
\mbox{}_{2}F_1 \left( \e, \e, 1+ \e, {1 \over 1 - a t }\right)
\right]
\ . 
\nonumber \\
\eeqa
Finally,  one can show that  \cite{dnizic}
\beq
F^{\rm 2me} (s,t,P^2, Q^2) = I(a,s) + I(a,t) - I(a,P^2) - I(a,Q^2)\ ,
\eeq
where
\beq
 I(a,s) = -\frac{c_\Gamma}{\epsilon^2}\Big(\frac{-s}{\mu^2}\Big)^{-\epsilon} +
  (-a\mu^2)^\epsilon c_\Gamma \sum_{n=2}^\infty
 \epsilon^{n-2}\Big( I_n(as) + \zeta_n \Big)\ ,
\eeq
with
\beq
I_n(x) = \frac{(-1)^n}{(n-1)!} \int_0^1 \frac{dz}{z}\ \log^{n-1}(1-z+xz)\ .
\eeq

%%%%%%%%%%%%%%%%%%%%%%%%%%%%%%%%%%%%%%%%%%%%%%%%%%%%%%%%%%%%%%%%%%%


\newpage
\begin{thebibliography}{99}

%%%%%%%%%%%%%%%%%%%%%%%%%%%%%%%%%%%%%%%%%%%%%%%%%%%%%%%%%%%%%%%%%%%



\bibitem{witten}
E.~Witten,
{\it Perturbative gauge theory as a string theory in twistor space,}
Commun.\ Math.\ Phys.\  {\bf 252}, 189 (2004),
{\tt hep-th/0312171}.
%%CITATION = HEP-TH 0312171;%%


\bibitem{csw}
F.~Cachazo, P.~Svr\v{c}ek  and E.~Witten,
{\it MHV vertices and tree amplitudes in gauge theory,}
JHEP {\bf 0409}, 006 (2004),
{\tt hep-th/0403047}.
%%CITATION = HEP-TH 0403047;%%



\bibitem{bible}
R.~J.~Eden, P.~V.~Landshoff, D.~I.~Olive and
J.~C.~Polkinghorne, {\it The Analytic S-Matrix},
Cambridge University Press, 1966.


\bibitem{bst}
A.~Brandhuber, B.~Spence and G.~Travaglini,
{\it One-loop gauge theory amplitudes in N = 4 super Yang-Mills from MHV
vertices,}
Nucl.\ Phys.\ B {\bf 706}, 150 (2005),
{\tt hep-th/0407214}.
%%CITATION = HEP-TH 0407214;%%



\bibitem{bdk1}
Z.~Bern, L.~J.~Dixon, D.~C.~Dunbar and D.~A.~Kosower,
{\it One loop n point gauge theory amplitudes, unitarity and collinear limits,}
Nucl.\ Phys.\ B {\bf 425} (1994) 217,
{\tt hep-ph/9403226}.
%%CITATION = HEP-PH 9403226;%%





\bibitem{Zhu}
C.~J.~Zhu,
{\it The googly amplitudes in gauge theory,}
JHEP {\bf 0404}, 032 (2004),
{\tt hep-th/0403115}.
%%CITATION = HEP-TH 0403115;%%


\bibitem{Georgiou:2004wu}
G.~Georgiou and V.~V.~Khoze,
{\it Tree amplitudes in gauge theory as scalar MHV diagrams,}
JHEP {\bf 0405}, 070 (2004),
{\tt hep-th/0404072}.
%%CITATION = HEP-TH 0404072;%%

\bibitem{WuZhutwo}
J.~B.~Wu and C.~J.~Zhu,
{\it MHV vertices and scattering amplitudes in gauge theory,}
JHEP {\bf 0407}, 032 (2004)
{\tt hep-th/0406085}.
%%CITATION = HEP-TH 0406085;%%


\bibitem{WuZhuthree}
J.~B.~Wu and C.~J.~Zhu,
{\it MHV vertices and fermionic scattering amplitudes in gauge theory with quarks
and gluinos,}
JHEP {\bf 0409}, 063 (2004)
{\tt hep-th/0406146}.
%%CITATION = HEP-TH 0406146;%%



\bibitem{Bena:2004ry}
  I.~Bena, Z.~Bern and D.~A.~Kosower,
  {\it Twistor-space recursive formulation of gauge theory amplitudes,}
  Phys.\ Rev.\ D {\bf 71}, 045008 (2005),
  {\tt hep-th/0406133}.
  %%CITATION = HEP-TH 0406133;%%



%\bibitem{dk}
%D.~Kosower, {\it Next-to-Maximal Helicity Violating Amplitudes in
%Gauge Theory},
%Phys.\ Rev.\  D {\bf 71} (2005) 045007, 
 %{\tt hep-th/0406175}.

\bibitem{ggk}
G.~Georgiou, E.~W.~N.~Glover and V.~V.~Khoze,
{\it Non-MHV tree amplitudes in gauge theory,}
JHEP {\bf 0407}, 048 (2004)
{\tt hep-th/0407027}.
%%CITATION = HEP-TH 0407027;%%



\bibitem{Ozeren:2005mp}
  K.~J.~Ozeren and W.~J.~Stirling,
  {\it MHV techniques for QED processes,}
JHEP {\bf 0511} (2005) 016, 
  {\tt hep-th/0509063}.
  %%CITATION = HEP-TH 0509063;%%




\bibitem{bcfw}
R.~Britto, F.~Cachazo, B.~Feng and E.~Witten,
{\it Direct proof of tree-level recursion relation in Yang-Mills theory,}
Phys.\ Rev.\ Lett.\  {\bf 94} (2005) 181602,
{\tt hep-th/0501052}.
%%CITATION = HEP-TH 0501052;%%


\bibitem{bcf}
R.~Britto, F.~Cachazo and B.~Feng,
{\it New recursion relations for tree amplitudes of gluons,}
Nucl.\ Phys.\ B {\bf 715}, 499 (2005)
{\tt hep-th/0412308}.
%%CITATION = HEP-TH 0412308;%%\\



\bibitem{bdk-rational1}
Z.~Bern, L.~J.~Dixon and D.~A.~Kosower,
{\it On-shell recurrence relations for one-loop QCD amplitudes,}
 Phys.\ Rev.\  D {\bf 71} (2005) 105013, {\tt hep-th/0501240}.

\bibitem{bdk-rational2}
Z.~Bern, L.~J.~Dixon and D.~A.~Kosower,
{\it The last of the finite loop amplitudes in QCD,}
Phys.\ Rev.\  D {\bf 72} (2005) 125003, 
{\tt hep-ph/0505055}.
%%CITATION = HEP-PH 0505055;%%


\bibitem{bdk-rational3}
Z.~Bern, L.~J.~Dixon and D.~A.~Kosower,
{\it Bootstrapping Multi-Parton Loop Amplitudes in QCD,}
Phys.\ Rev.\  D {\bf 73} (2006) 065013, 
{\tt hep-ph/0507005}.


\bibitem{coeffrec}
Z.~Bern, N.~E.~J.~Bjerrum-Bohr, D.~C.~Dunbar and H.~Ita,
{\it Recursive Calculation of
One-Loop QCD Integral Coefficients,}
JHEP {\bf 0511} (2005) 027, 
{\tt hep-ph/0507019}.



\bibitem{snvp}
S.~D.~Badger, E.~W.~N.~Glover, V.~V.~Khoze and P.~Svr\v{c}ek,
{\it Recursion relations for gauge theory amplitudes with massive particles,}
JHEP {\bf 0507} (2005) 025, 
{\tt  hep-th/0504159}.
%%CITATION = HEP-TH 0504159;%%


\bibitem{Badger:2005jv}
  S.~D.~Badger, E.~W.~N.~Glover and V.~V.~Khoze,
  {\it Recursion relations for gauge theory amplitudes with massive vector bosons
  and fermions,}
  JHEP {\bf 0601} (2006) 066, 
  {\tt hep-th/0507161}.
  %%CITATION = HEP-TH 0507161;%%



\bibitem{lnv}
L.~J.~Dixon, E.~W.~N.~Glover and V.~V.~Khoze,
{\it MHV rules for Higgs plus multi-gluon amplitudes,}
JHEP {\bf 0412} (2004) 015,
{\tt hep-th/0411092}.
%%CITATION = HEP-TH 0411092;%%


\bibitem{risager}
  K.~Risager,
  {\it A direct proof of the CSW rules,}
  JHEP {\bf 0512} (2005) 003, 
  {\tt hep-th/0508206}.
  %%CITATION = HEP-TH 0508206;%%




\bibitem{bbst3}
J.~Bedford, A.~Brandhuber, B.~Spence and G.~Travaglini,
{\it A recursion relation for gravity amplitudes,}
Nucl.\ Phys.\ B {\bf 721} (2005) 98
{\tt hep-th/0502146}.
%%CITATION = HEP-TH 0502146;%%

\bibitem{cs}
F.~Cachazo and P.~Svr\v{c}ek,
{\it Tree level recursion relations in general relativity,}
{\tt hep-th/0502160}.
%%CITATION = HEP-TH 0502160;%%



\bibitem{swan}
N.~E.~J.~Bjerrum-Bohr, D.~C.~Dunbar,
H.~Ita, W.~B.~Perkins and K.~Risager,
{\it MHV-vertices for gravity amplitudes,}
JHEP {\bf 0601} (2006) 009, 
{\tt hep-th/0509016}.
  %%CITATION = HEP-TH 0509016;%%


\bibitem{bbst}
J.~Bedford, A.~Brandhuber, B.~Spence and G.~Travaglini,
{\it A twistor approach to one-loop amplitudes in N = 1 supersymmetric Yang-Mills
theory,}
Nucl.\ Phys.\ B {\bf 706}, 100 (2005),
{\tt hep-th/0410280}.
%%CITATION = HEP-TH 0410280;%%


\bibitem{qr}
C.~Quigley and M.~Rozali,
{\it One-loop MHV amplitudes in supersymmetric gauge theories,}
JHEP {\bf 0501} (2005) 053
{\tt hep-th/0410278}.
%%CITATION = HEP-TH 0410278;%%


\bibitem{bbst2}
J.~Bedford, A.~Brandhuber, B.~Spence and G.~Travaglini,
{\it Non-supersymmetric loop amplitudes and MHV vertices,}
Nucl.\ Phys.\ B {\bf 712}, 59 (2005)
{\tt hep-th/0412108}.
%%CITATION = HEP-TH 0412108;%%


\bibitem{Bern:1994cg}
Z.~Bern, L.~J.~Dixon, D.~C.~Dunbar and D.~A.~Kosower, {\it Fusing
gauge theory tree amplitudes into loop amplitudes,} Nucl.\ Phys.\
B {\bf 435} (1995) 59, {\tt hep-ph/9409265}.

\bibitem{bdk9302280}
Z.~Bern, L.~J.~Dixon and D.~A.~Kosower, {\it One-Loop Corrections to Five-Gluon
amplitudes,} Phys. Rev. Lett. {\bf 70} (1993) 2677-2680,
{\tt hep-ph/9302280}.



\bibitem{F1}
  R.~P.~Feynman,
  {\it Quantum Theory Of Gravitation,}
  Acta Phys.\ Polon.\  {\bf 24} (1963) 697.
  %%CITATION = APPOA,24,697;%%




\bibitem{F2}
  R.~P.~Feynman,
  {\it Closed Loop And Tree Diagrams,}
 in J.~R.~Klauder, {\it Magic Without Magic},
 San Francisco 1972, 355-375; in  Brown, L.~M.~(ed.):
 {\it Selected papers of Richard Feynman,} 867-887



\bibitem{F3}
  R.~P.~Feynman,
  {\it Problems In Quantizing The Gravitational Field, And The Massless
  Yang-Mills Field,}
 In J.~R.~Klauder, {\it Magic Without Magic}, San Francisco 1972, 377-408;
 in Brown, L.~M.~(ed.):
{\it Selected papers of Richard Feynman}, 888-919.




\bibitem{csw2}
F.~Cachazo, P.~Svr\v{c}ek and E.~Witten,
{\it Twistor space structure of one-loop amplitudes in gauge theory,}
JHEP {\bf 0410} (2004) 074,
{\tt hep-th/0406177}.
%%CITATION = HEP-TH 0406177;%%


%\cite{Bern:1995ix}
\bibitem{Bern:1995ix}
Z.~Bern and G.~Chalmers,
{\it Factorization in one loop gauge theory,}
Nucl.\ Phys.\ B {\bf 447}, 465 (1995),
{\tt  hep-ph/9503236}.
%%CITATION = HEP-PH 9503236;%%





\bibitem{david}
D.~A.~Kosower,
{\it All-order collinear behavior in gauge theories,}
Nucl.\ Phys.\ B {\bf 552} (1999) 319,
{\tt hep-ph/9901201}.
%%CITATION = HEP-PH 9901201;%%

\bibitem{ku}
D.~A.~Kosower and P.~Uwer,
{\it One-loop splitting amplitudes in gauge theory,}
Nucl.\ Phys.\ B {\bf 563} (1999) 477,
{\tt hep-ph/9903515}.
%%CITATION = HEP-PH 9903515;%%



\bibitem{vittorio}
Z.~Bern, V.~Del Duca, W.~B.~Kilgore and C.~R.~Schmidt,
{\it The infrared behavior of one-loop {QCD} amplitudes at
next-to-next-to-leading order,}
Phys.\ Rev.\ D {\bf 60} (1999) 116001,
{\tt hep-ph/9903516}.
%%CITATION = HEP-PH 9903516;%%



\bibitem{vittorio2}
Z.~Bern, V.~Del Duca and C.~R.~Schmidt,
{\it The infrared behavior of one-loop gluon amplitudes at
next-to-next-to-leading order,}
Phys.\ Lett.\ B {\bf 445} (1998) 168,
{\tt hep-ph/9810409}.
%%CITATION = HEP-PH 9810409;%%





\bibitem{csw3}
F.~Cachazo, P.~Svr\v{c}ek and E.~Witten,
{\it Gauge theory amplitudes in twistor space and holomorphic anomaly,}
JHEP {\bf 0410} (2004) 077,
{\tt hep-th/0409245}.
%%CITATION = HEP-TH 0409245;%%


\bibitem{lits}
I.~Bena, Z.~Bern, D.~A.~Kosower and R.~Roiban,
{\it Loops in twistor space,}
Phys.\ Rev.\ D {\bf 71} (2005) 106010,
{\tt hep-th/0410054}.
%%CITATION = HEP-TH 0410054;%%


\bibitem{Cachazo:2004dr}
F.~Cachazo,
{\it Holomorphic anomaly of unitarity cuts and one-loop gauge theory amplitudes,}
{\tt hep-th/0410077}.
%%CITATION = HEP-TH 0410077;%%


\bibitem{Britto:2004nj}
R.~Britto, F.~Cachazo and B.~Feng,
{\it Computing one-loop amplitudes from the holomorphic anomaly of unitarity
cuts,}
Phys.\ Rev.\ D {\bf 71} (2005) 025012,
{\tt hep-th/0410179}.
%%CITATION = HEP-TH 0410179;%%


%\cite{Bern:2004ky}
\bibitem{Bern:2004ky}
Z.~Bern, V.~Del Duca, L.~J.~Dixon and D.~A.~Kosower,
{\it All non-maximally-helicity-violating one-loop seven-gluon amplitudes in N = 4
super-Yang-Mills theory,}
Phys.\ Rev.\ D {\bf 71} (2005) 045006,
{\tt hep-th/0410224}.
%%CITATION = HEP-TH 0410224;%%


%\cite{Bidder:2004tx}
\bibitem{Bidder:2004tx}
S.~J.~Bidder, N.~E.~J.~Bjerrum-Bohr, L.~J.~Dixon and D.~C.~Dunbar,
{\it N = 1 supersymmetric one-loop amplitudes and the holomorphic anomaly of
unitarity cuts,}
Phys.\ Lett.\ B {\bf 606} (2005) 189,
{\tt hep-th/0410296}.
%%CITATION = HEP-TH 0410296;%%



\bibitem{bcf-gen}
R.~Britto, F.~Cachazo and B.~Feng,
{\it Generalized unitarity and one-loop amplitudes in N = 4 super-Yang-Mills,}
Nucl.\ Phys.\  B {\bf 725}, 275 (2005), 
{\tt hep-th/0412103}.
%%CITATION = HEP-TH 0412103;%%


\bibitem{Bern:2004ba}
Z.~Bern, D.~Forde, D.~A.~Kosower and P.~Mastrolia,
{\it Twistor-inspired construction of electroweak vector boson currents,}
Phys.\ Rev.\  D {\bf 72} (2005) 025006, 
{\tt hep-ph/0412167}.
%%CITATION = HEP-PH 0412167;%%


\bibitem{bdk-december}
Z.~Bern, L.~J.~Dixon and D.~A.~Kosower,
{\it All next-to-maximally helicity-violating one-loop gluon amplitudes in N = 4
super-Yang-Mills theory,}
Phys.\ Rev.\  D {\bf 72}, 045014 (2005), 
{\tt hep-th/0412210}.
%%CITATION = HEP-TH 0412210;%%




\bibitem{BBDP}
S.~J.~Bidder, N.~E.~J.~Bjerrum-Bohr, D.~C.~Dunbar and W.~B.~Perkins,
{\it One-loop gluon scattering amplitudes in theories with $N < 4$
supersymmetries,}
Phys.\ Lett.\ B {\bf 612} (2005) 75,
{\tt hep-th/0502028}.
%%CITATION = HEP-TH 0502028;%%


\bibitem{BBCF}
R.~Britto, E.~Buchbinder, F.~Cachazo and B.~Feng,
{\it One-loop amplitudes of gluons in SQCD,}
 Phys.\ Rev.\  D {\bf 72}, 065012 (2005), 
{\tt hep-ph/0503132}.
%%CITATION = HEP-PH 0503132;%%


\bibitem{lance}
L.~J.~Dixon,
{\it Calculating scattering amplitudes efficiently,}
{\tt hep-ph/9601359}.
%%CITATION = HEP-PH 9601359;%%



\bibitem{bgkm1}
T.~G.~Birthwright, E.~W.~N.~Glover, V.~V.~Khoze and P.~Marquard,
{\it Multi-gluon collinear limits from MHV diagrams,}
JHEP {\bf 0505} (2005) 013, 
{\tt hep-ph/0503063}.
%%CITATION = HEP-PH 0503063;%%

\bibitem{bgkm2}
T.~G.~Birthwright, E.~W.~N.~Glover, V.~V.~Khoze and P.~Marquard,
{\it Collinear limits in QCD from MHV rules,}
JHEP {\bf 0507} (2005) 068, 
{\tt hep-ph/0505219}.
%%CITATION = HEP-PH 0505219;%%



\bibitem{Kunszt:1993sd}
Z.~Kunszt, A.~Signer and Z.~Trocsanyi,
{\it One loop helicity amplitudes for all 2 $\to$ 2 processes in QCD and N=1
supersymmetric Yang-Mills theory,}
Nucl.\ Phys.\ B {\bf 411} (1994) 397,
{\tt hep-ph/9305239}.
%%CITATION = HEP-PH 9305239;%%



\bibitem{Bern:1991aq}
Z.~Bern and D.~A.~Kosower,
{\it The Computation of loop amplitudes in gauge theories,}
Nucl.\ Phys.\ B {\bf 379}, 451 (1992).
%%CITATION = NUPHA,B379,451;%%


\bibitem{ES}
R.~K.~Ellis and J.~C.~Sexton,
{\it QCD Radiative Corrections To Parton Parton Scattering,}
Nucl.\ Phys.\ B {\bf 269}, 445 (1986).
%%CITATION = NUPHA,B269,445;%%

\bibitem{Bern:1993mq}
Z.~Bern, L.~J.~Dixon and D.~A.~Kosower,
{\it One loop corrections to five gluon amplitudes,}
Phys.\ Rev.\ Lett.\  {\bf 70}, 2677 (1993),
{\tt hep-ph/9302280}.
%%CITATION = HEP-PH 9302280;%%

%\cite{Kunszt:1994tq}
\bibitem{KST}
Z.~Kunszt, A.~Signer and Z.~Trocsanyi,
{\it One loop radiative corrections to the helicity amplitudes of QCD processes
involving four quarks and one gluon,}
Phys.\ Lett.\ B {\bf 336}, 529 (1994), 
{\tt hep-ph/9405386}.
%%CITATION = HEP-PH 9405386;%%

%\cite{Bern:1994fz}
\bibitem{Bern:1994fz}
Z.~Bern, L.~J.~Dixon and D.~A.~Kosower,
{\it One loop corrections to two quark three gluon amplitudes,}
Nucl.\ Phys.\ B {\bf 437} (1995) 259,
{\tt hep-ph/9409393}.
%%CITATION = HEP-PH 9409393;%%





\bibitem{BMST}
A.~Brandhuber, S.~McNamara, B.~Spence and G.~Travaglini,
{\it Loop amplitudes in pure Yang-Mills
from generalised unitarity,} 
JHEP {\bf 0510}, 011 (2005)
 {\tt hep-th/0506068}.
%%CITATION = JHEPA,0510,011;%%


\bibitem{Huang:2005ve}
Y.~t.~Huang,
{\it N = 4 SYM NMHV loop amplitude in superspace,}
Phys.\ Lett.\  B {\bf 631} (2005) 177, 
{\tt hep-th/0507117}.
%%CITATION = HEP-TH 0507117;%%

\bibitem{Risager:2005ke}
K.~Risager, S.~J.~Bidder and W.~B.~Perkins,
{\it One-loop NMHV amplitudes involving gluinos and scalars in N = 4 gauge
theory,} JHEP {\bf 0510} (2005) 003, 
{\tt hep-th/0507170}.
%%CITATION = HEP-TH 0507170;%%


\bibitem{Forde:2005ue}
D.~Forde and D.~A.~Kosower,
{\it All-multiplicity amplitudes with massive scalars,}
Phys.\ Rev.\  D {\bf 73} (2006) 065007, 
{\tt hep-th/0507292}.
%%CITATION = HEP-TH 0507292;%%

%\cite{Forde:2005hh}
\bibitem{Forde:2005hh}
D.~Forde and D.~A.~Kosower,
{\it All-multiplicity one-loop corrections to MHV amplitudes in QCD,}
Phys.\ Rev.\  D {\bf 73} (2006) 061701, {\tt hep-ph/0509358}.
%%CITATION = HEP-PH 0509358;%%


\bibitem{babis1}
C.~Anastasiou, Z.~Bern, L.~J.~Dixon and D.~A.~Kosower,
{\it Planar amplitudes in maximally supersymmetric Yang-Mills theory,}
Phys.\ Rev.\ Lett.\  {\bf 91} (2003) 251602, 
{\tt hep-th/0309040}.
%%CITATION = HEP-TH 0309040;%%


\bibitem{babis2}
C.~Anastasiou,  Z.~Bern, L.~J.~Dixon, and D.~A.~Kosower,
{\it Cross-order relations in N = 4 supersymmetric gauge theories,}
{\tt hep-th/0402053}.
%%CITATION = HEP-TH 0402053;%%

\bibitem{bds}
Z.~Bern, L.~J.~Dixon and V.~A.~Smirnov,
{\it Iteration of planar amplitudes in maximally supersymmetric Yang-Mills theory
at three loops and beyond,}
Phys.\ Rev.\  D {\bf 72} (2005) 085001, 
{\tt hep-th/0505205}.
%%CITATION = HEP-TH 0505205;%%

\bibitem{Buc-Cac}
E.~I.~Buchbinder and F.~Cachazo,
{\it Two-loop amplitudes of gluons and octa-cuts in N = 4 super Yang-Mills,}
JHEP {\bf 0511} (2005) 036, 
{\tt hep-th/0506126}.
%%CITATION = HEP-TH 0506126;%%



\bibitem{Landau:1959fi}
L.~D.~Landau,
{\it On Analytic Properties Of Vertex Parts In Quantum Field Theory,}
Nucl.\ Phys.\  {\bf 13} (1959) 181.
%%CITATION = NUPHA,13,181;%%


\bibitem{Cutkosky:1960sp}
R.~E.~Cutkosky,
{\it Singularities And Discontinuities Of Feynman Amplitudes,}
J.\ Math.\ Phys.\  {\bf 1} (1960) 429.
%%CITATION = JMAPA,1,429;%%





\bibitem{zuzu'}
  C.~Itzykson and J.~B.~Zuber,
  {\it Quantum Field Theory,} McGraw-Hill (1985).



\bibitem{Kosower:2004yz}
  D.~A.~Kosower,
  {\it Next-to-maximal helicity violating amplitudes in gauge theory,}
  Phys.\ Rev.\ D {\bf 71}, 045007 (2005),
  {\tt hep-th/0406175}.
  %%CITATION = HEP-TH 0406175;%%





\bibitem{Nair}
  V.~P.~Nair,
  {\it A Current Algebra For Some Gauge Theory Amplitudes,}
  Phys.\ Lett.\ B {\bf 214} (1988) 215.
  %%CITATION = PHLTA,B214,215;%%



\bibitem{dnizic}
G.~Duplancic and B.~Nizic,
{\it Dimensionally regulated one-loop box
scalar integrals with massless internal lines,}
Eur. Phys. J. {\bf C20} (2001) 357-370,
{\tt hep-ph/0006249}.
%%CITATION = HEP-PH 0006249;%%


\bibitem{ultimo}
Z.~Bern, L.~J.~Dixon and D.~A.~Kosower,
{\it Dimensionally regulated pentagon integrals,}
Nucl.\ Phys.\ B {\bf 412} (1994) 751, 
{\tt hep-ph/9306240}.
%%CITATION = HEP-PH 9306240;%%


\end{thebibliography}
\end{document}